\documentclass[aps,reprint,eqsecnum,pre,longbibliography]{revtex4-1}
\usepackage[dvips]{graphicx}
\usepackage{color}
\usepackage{amssymb,amsmath}
\usepackage{bm}
\usepackage{pstricks,pst-grad}
\usepackage{amsbsy}
\usepackage{epic}
\usepackage[normalem]{ulem}
\usepackage{hyperref}
\usepackage[hyphenbreaks]{breakurl}

\def\bal#1\eal{\begin{align}#1\end{align}}
\newcommand\beq{\begin{equation}}
\newcommand\eeq{\end{equation}}
\newcommand\beqa{\begin{eqnarray}}
\newcommand\eeqa{\end{eqnarray}}
\newcommand{\nn}{\nonumber\\}

\newcommand{\ed}{\end{document}}

\newcommand{\ma}{m_i}

\newcommand{\mb}{m_j}

\newcommand{\cca}{\mathbf{v}_i}

\newcommand{\ca}{{v}_i}

\newcommand{\ccb}{\mathbf{v}_j}

\newcommand{\cb}{{v}_j}

\newcommand{\JJ}{\mathbf{Q}_{ij}}
\newcommand{\JJw}{{\mathbf{Q}}_{ij}^-}

\newcommand{\kk}{\widehat{\bm{\sigma}}}

\newcommand{\wwa}{\bm{\omega}_i}

\newcommand{\wwb}{\bm{\omega}_j}

\newcommand{\wa}{{\omega}_i}

\newcommand{\wb}{{\omega}_j}

\newcommand{\Ia}{I_i}

\newcommand{\Ib}{I_j}

\newcommand{\da}{\sigma_i}

\newcommand{\db}{\sigma_j}

\newcommand{\ds}{\sigma}

\newcommand{\dab}{\sigma_{ij}}

\newcommand{\x}{\times}

\newcommand{\gh}{\mathbf{v}_{ij}}

\newcommand{\g}{{v}_{ij}}

\newcommand{\een}{\alpha_{ij}}

\newcommand{\esn}{\alpha}

\newcommand{\eet}{\beta_{ij}}

\newcommand{\est}{\beta}

\newcommand{\enn}{\overline{\alpha}_{ij}}

\newcommand{\ett}{\overline{\beta}_{ij}}

\newcommand{\mab}{m_{ij}}

\newcommand{\qab}{\kappa_{ij}}

\newcommand{\qa}{\kappa_{i}}

\newcommand{\qb}{\kappa_{j}}

\newcommand{\q}{\kappa}

\newcommand{\fa}{f_{i}}

\newcommand{\fb}{f_{j}}

\newcommand{\fab}{f_{ij}^{(2)}}

\newcommand{\ffab}{\bar{f}_{ij}^{(2)}}

\newcommand{\fat}{f_{i}^\text{tr}}

\newcommand{\far}{f_{i}^\text{rot}}

\newcommand{\fbr}{f_{j}^\text{rot}}

\newcommand{\Tat}{T_{i}^\text{tr}}

\newcommand{\Tbt}{T_{j}^\text{tr}}

\newcommand{\Tt}{T^\text{tr}}

\newcommand{\Tar}{T_{i}^\text{rot}}

\newcommand{\Tbr}{T_{j}^\text{rot}}

\newcommand{\Tr}{T^\text{rot}}

\newcommand{\Qab}{J_{ij}}

\newcommand{\Iab}{\mathcal{J}_{ij}}

\newcommand{\na}{n_i}

\newcommand{\nb}{n_j}

\newcommand{\zabt}{\xi_{ij}^\text{tr}}

\newcommand{\zt}{\xi^\text{tr}}

\newcommand{\zabr}{\xi_{ij}^\text{rot}}

\newcommand{\zr}{\xi^\text{rot}}

\newcommand{\al}{i}

\newcommand{\be}{j}

\newcommand{\tr}{\text{tr}}

\newcommand{\rot}{\text{rot}}

\newcommand{\NN}{s}

\newcommand{\wwwa}{\mathbf{w}_i}
\newcommand{\wwwb}{\mathbf{w}_j}
\newcommand{\wwwab}{\mathbf{w}_{ij}}

\newcommand{\Sab}{{S}_{ij}}
\newcommand{\chiab}{{\chi}_{ij}}

\newcommand{\uk}{\widehat{\mathbf{k}}}
\newcommand{\llangle}{\langle\!\langle}
\newcommand{\rrangle}{\rangle\!\rangle}

\newcommand{\xx}{\mathbf{c}}

\newcommand{\ancho}{0.8}
\newcommand{\anchotwo}{1.7}

\begin{document}



\title{Interplay between polydispersity, inelasticity, and roughness in the freely cooling regime of hard-disk granular gases}

\author{Andr\'es Santos}
\email{andres@unex.es}
\homepage{http://www.unex.es/eweb/fisteor/andres/}
\affiliation{Departamento de F\'{\i}sica and Instituto de Computaci\'on Cient\'ifica Avanzada (ICCAEx), Universidad de Extremadura,
E-06071 Badajoz, Spain}

\date{\today}

\begin{abstract}
A polydisperse granular gas made of inelastic and rough hard disks is considered. Focus is laid on the kinetic-theory derivation of the partial energy production rates and  the total cooling rate as functions of the partial densities and temperatures (both translational and rotational) and of the parameters of the mixture (masses, diameters, moments of inertia, and mutual coefficients of normal and tangential restitution). The results are applied to the homogeneous cooling state of the system and the associated nonequipartition of energy among the different components and degrees of freedom.
It is found that disks typically present a stronger rotational-translational nonequipartition but a weaker component-component nonequipartition than spheres.
A noteworthy ``mimicry'' effect is unveiled, according to which a polydisperse gas of disks having common values of the coefficient of restitution and of the reduced moment of inertia can be made indistinguishable from  a monodisperse gas in what concerns the degree of rotational-translational energy nonequipartition. This effect requires the mass of a disk of component $i$ to be approximately proportional to $2\sigma_i+\langle\sigma\rangle$, where $\sigma_i$ is the diameter of the disk and $\langle\sigma\rangle$ is the mean diameter.

\end{abstract}

\maketitle

\section{Introduction}

The minimal model to describe the dynamical properties of a granular fluid consists of a collection of identical, smooth  hard disks (in two-dimensional geometry) or spheres (in the three-dimensional case). Particles dissipate kinetic energy via binary collisions and this is characterized in the minimal model by means of a constant coefficient of normal restitution.  While this simple model  captures most of the basic properties of granular flows \cite{D00,OK00,PL01,G03,K04,BP04,AT06,RN08}, it can be made more realistic, for instance,  by assuming that the coefficient of normal restitution depends on  the impact velocity \cite{BP04,BSSP04,DF17}, taking into account the presence of an interstitial fluid \cite{XVKL09}, considering  non-spherical particles \cite{HZMP09},  introducing the effect of surface friction in collisions, or  accounting for a multicomponent character of the granular fluid.

In particular, there exists a vast literature about polydisperse systems of smooth disks or spheres \cite{JM89,GD99b,HQL01,JY02,MG02b,BT02a,DHGD02,GD02,%
BRM05,SGNT06,GDH07,GHD07,G08,G08b,UKAZ09}, as well as about friction (or roughness) in monodisperse systems
\cite{JR85a,LS87,C89,L91,LB94,GS95,L95,L96,ZTPSH98,HZ97,ML98,LHMZ98,HHZ00,AHZ01,MHN02,CLH02,JZ02,PZMZ02,MSS04,HCZHL05,GNB05,Z06,BPKZ07,GA08,KBPZ09,K10a,%
S11a,SKS11,SK12,MDHEH13,VSK14,VSK14b,KSG14,RA14,VS15,FH17,SP17,DF17,GSK18}.
On the other hand, much fewer works have dealt with multicomponent gases of rough spheres \cite{VT04,PTV07,CP08,SKG10,S11b,VLSG17,VLSG17b}. This class of systems is especially relevant because of an inherent breakdown of energy equipartition, even in homogeneous and isotropic states (driven or undriven), as characterized by independent translational ($\Tat$) and rotational ($\Tar$) temperatures associated with each component $i$. The rate of change of the  translational mean kinetic energy of particles of component $i$ due to collisions with particles of component $j$ defines the energy production rate $\zabt$. A similar energy production rate $\zabr$ measures the rate of change of the  rotational mean kinetic energy.

By means of kinetic-theory tools, the energy production rates $\zabt$ and $\zabr$ for (three-dimensional) hard spheres were obtained in Ref.\ \cite{SKG10} as functions of $\Tat$, $\Tbt$, $\Tar$, $\Tbr$, and of the mechanical parameters (masses, diameters, moments of inertia, and coefficients of normal and tangential restitution) of each pair $ij$. Those expressions were derived by assuming collisional molecular chaos, statistical independence between the translational and angular velocities, and a Maxwellian form for the translational velocity distribution function. The application of the results to the homogeneous cooling state (HCS) of a tracer particle immersed in a granular gas of inelastic and rough hard spheres shows a very good agreement with computer simulations \cite{VLSG17,VLSG17b}.

From the experimental point of view, however, most of the setup geometries are two-dimensional \cite{OU98,RM00,FM02,SM08,DPLD09,GBG09,TMHS09,ND12,APGZM13,GBM15,SSP16,SP17}. Moreover, while capturing most of the physics of the problems at hand,
two-dimensional computer simulations are much easier to carry out and interpret than three-dimensional ones. Hence, the extension of the analysis carried out in Ref.\ \cite{SKG10} to  multicomponent hard disks has undoubtedly a practical interest beyond its  added academic value. In contrast to what happens for smooth, spinless particles, where an unambiguous kinetic-theory treatment of $d$-dimensional hard spheres is possible \cite{VGS07}, the existence of angular motion due to surface friction or roughness establishes a neat separation between the cases of  spheres and  disks. Whereas both classes of particles are embedded in a common three-dimensional space, spinning spheres have three translational plus three rotational degrees of freedom, but spinning disks on a plane have two translational and only one rotational degrees of freedom.

By following steps similar to those followed in Ref.\ \cite{SKG10}, the energy production rates $\zabt$ and $\zabr$ are derived in this paper for a multicomponent gas made of inelastic and rough disks. The results are subsequently applied to the HCS and illustrated for monodisperse and bidisperse gases. An interesting \emph{mimicry} effect is also analyzed. According to this effect, the HCS of a polydisperse gas of disks having common values of the coefficient of restitution and of the reduced moment of inertia can be indistinguishable from that of a monodisperse gas in what concerns the rotational-translational temperature ratio. It is shown here that the condition for this mimicry effect is that the mass $m_i$ of each component $i$ must be approximately proportional to $2\sigma_i+\langle\sigma\rangle$, where $\sigma_i$ is the diameter of a disk of component $i$ and $\langle\sigma\rangle$ is the mean diameter.

The organization of this paper is as follows. Section \ref{sec2} describes the collision rules, which are then  used in Sec.\ \ref{sec3} to express the collisional rates of change in terms of two-body averages. Next, those averages are estimated by assuming molecular chaos, statistical independence between the translational and angular velocities, and a Maxwellian translational velocity distribution function. The energy production rates $\zabt$ and $\zabr$ are defined in Sec.\ \ref{sec5}, their explicit expressions being displayed in Table \ref{table:2}. Those results are applied to the HCS of monodisperse and bidisperse systems in Sec.\ \ref{sec6}. Section \ref{sec7} deals with the mimicry effect described above.
Finally, the paper ends with some concluding remarks in Sec.\ \ref{sec:conc}.

\section{Binary collisions. Coefficients of restitution}
\label{sec2}

\subsection{Collisional rules}

Let us consider an $\NN$-component granular gas of hard disks  (lying on the $xy$ plane). Disks of component $i$ have a mass $\ma$, a diameters $\da$, and a moment of inertia $\Ia= \frac{1}{4}\ma\da^2\qa$,
where the value of the dimensionless quantity $\qa$ depends on the mass distribution within the disk, running from the extreme values $\qa=0$ (mass concentrated on the center) to $\qa=1$ (mass concentrated on the perimeter). If the mass is uniformly distributed, then $\qa=\frac{1}{2}$.

\begin{figure}[tbp]
\centerline{\includegraphics[width=\ancho\columnwidth]{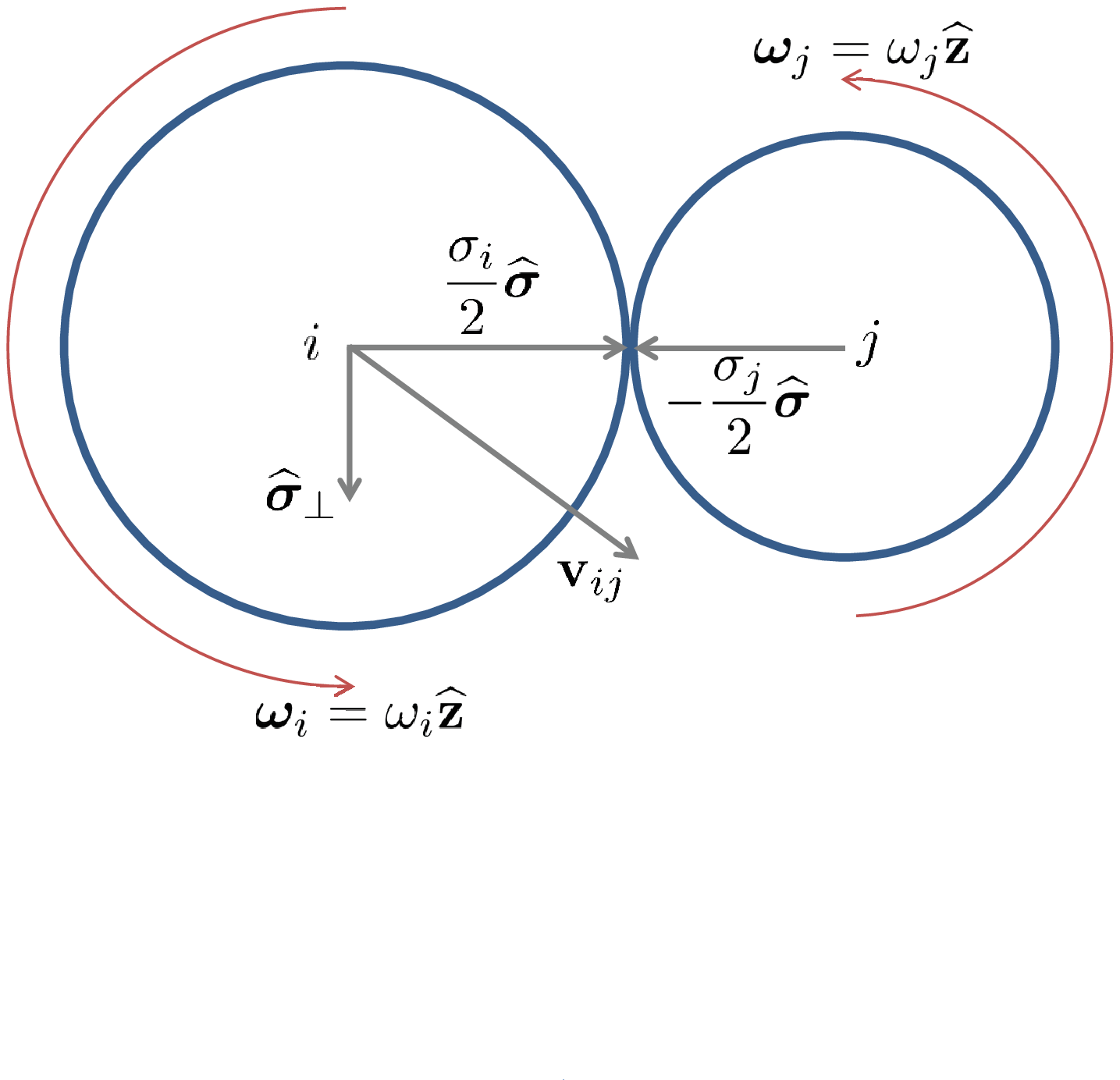}}
\caption{Sketch of the pre-collisional quantities of disks $i$ and $j$ in the frame of reference solidary with disk $j$.}
\label{sketch}
\end{figure}

Figure \ref{sketch} sketches a binary collision between two disks of components $i$ and $j$.
Let us denote by $\gh=\cca-\ccb$ the pre-collisional relative velocity of the center of mass of both disks, by $\wwa=\wa\widehat{\mathbf{z}}$ and $\wwb=\wb\widehat{\mathbf{z}}$ the respective pre-collisional angular velocities, by $\kk\equiv (\mathbf{r}_j-\mathbf{r}_i)/|\mathbf{r}_j-\mathbf{r}_i|$ the unit vector pointing from the center of $i$ to the center of $j$, and by $\kk_\perp=\kk\times\widehat{\mathbf{z}}=\widehat{\sigma}_y\widehat{\mathbf{x}}-\widehat{\sigma}_x\widehat{\mathbf{y}}$ its perpendicular unit vector. The velocities of  the points of the disks which
are in contact at the collision are
\begin{equation}
\wwwa=\cca-\frac{\da}{2}\wa\kk_\perp,\quad \wwwb=\ccb+\frac{\db}{2}\wb\kk_\perp,
\label{w}
\end{equation}
so that the corresponding relative velocity is
\begin{equation}
\wwwab=\gh-\Sab\kk_\perp,\quad \Sab\equiv \frac{\da}{2}\wa+ \frac{\db}{2}\wb.
\label{Sab}
\end{equation}

Post-collisional velocities will be denoted by primes.  The conservation of linear and angular momenta yields
\begin{subequations}
\label{conservation}
\begin{equation}
\ma\cca'+\mb\ccb'=\ma\cca+\mb\ccb,
\label{momentum}
\end{equation}
\begin{equation}
\label{ang_mom_1}
\Ia\wa'+\ma\frac{\da}{2}\cca'\cdot\kk_\perp=\Ia\wa+\ma\frac{\da}{2}\cca\cdot\kk_\perp,
\end{equation}
\begin{equation}
\label{ang_mom_2}
\Ib\wb'-\mb\frac{\db}{2}\ccb'\cdot\kk_\perp=\Ib\wb-\mb\frac{\db}{2}\ccb\cdot\kk_\perp.
\end{equation}
\end{subequations}
Angular momentum (with respect to the point of contact) is conserved for each particle separately because during a collision the forces act only at the point of contact and hence there is no torque with respect to that point \cite{Z06}.
Equations \eqref{conservation}  imply that
\begin{subequations}
\label{13b&14}
\begin{equation}
\cca'=\cca-\frac{1}{\ma}\JJ,\quad \ccb'=\ccb+\frac{1}{\mb}\JJ,
\label{13b}
\end{equation}
\begin{equation}
\wa'=\wa+\frac{\da}{2\Ia}\JJ\cdot\kk_\perp, \quad \wb'=\wb+\frac{\db}{2\Ib}\JJ\cdot\kk_\perp,
\label{14}
\end{equation}
\end{subequations}
where the (so-far) undetermined quantity $\JJ$ is the impulse exerted by particle $i$ on particle $j$.
Therefore, the post-collisional relative velocities are
\begin{subequations}
\label{wwwab'}
  \begin{equation}
\gh'=\gh-\frac{1}{\mab}\JJ,
\end{equation}
\begin{equation}
\label{wwwwab'}
\wwwab'=\wwwab-\frac{1}{\mab}\JJ-\frac{1}{\mab\qab}\left(\JJ\cdot\kk_\perp\right)\kk_\perp,
\end{equation}
\end{subequations}
where
\begin{equation}
\mab\equiv \frac{\ma\mb}{\ma+\mb},\quad \qab\equiv \qa\qb\frac{\ma+\mb}{\qa\ma+\qb\mb}
\label{21}
\end{equation}
are the reduced mass and a sort of reduced inertia-moment parameter, respectively.

The collisional rules can be closed by relating the normal (i.e., parallel to $\kk$) and tangential (i.e., parallel to $\kk_\perp$) components of the relative velocities $\wwwab$ and $\wwwab'$:
\begin{equation}
\wwwab'\cdot\kk=-\een \wwwab\cdot\kk,\quad  \wwwab'\cdot\kk_\perp=-\eet \wwwab\cdot\kk_\perp.
\label{restitution}
\end{equation}
Here, $\een$ and $\eet$ are the  \emph{constant} coefficients of normal and tangential restitution, respectively. While $\een$ ranges from $\een=0$ (perfectly inelastic particles) to $\een=1$ (perfectly elastic particles), the coefficient $\eet$ runs from $\eet=-1$ (perfectly smooth particles, i.e., no change in the tangential component of the relative velocity) to $\eet=1$ (perfectly rough particles, i.e., reversal of the tangential component).
The insertion of  Eq.\ \eqref{wwwwab'} into Eq.\ \eqref{restitution} yields
\begin{equation}
\label{JJ}
\frac{\JJ\cdot\kk}{\mab}=\enn \wwwab\cdot\kk,\quad \frac{\JJ\cdot\kk_\perp}{\mab}=\ett \wwwab\cdot\kk_\perp,
\end{equation}
with the introduction of the parameters
\begin{equation}
\enn\equiv1+\een,\quad\ett\equiv\frac{\qab}{1+\qab}\left(1+\eet\right).
\label{20}
\end{equation}
Therefore, with the help of Eqs.\ \eqref{Sab} and \eqref{JJ}, the impulse $\JJ$ is expressed in terms of the pre-collisional velocities and the unit vector $\kk$ as
\begin{equation}
\frac{\JJ}{\mab}=\enn (\gh\cdot\kk)\kk+\ett\left(\gh\cdot\kk_\perp-\Sab\right)\kk_\perp.
\label{15}
\end{equation}
This, together with Eqs.\ \eqref{13b&14}, closes the collision rules $(\mathbf{v}_i,\omega_i;\mathbf{v}_j,\omega_j)\stackrel{\kk}{\to}(\mathbf{v}_i',\omega_i';\mathbf{v}_j',\omega_j')$.
Note that  one has $\ett=0$ in the special case of perfectly smooth disks ($\eet=-1$), so that $\JJ\cdot\kk_\perp={0}$ in that case and, according to Eq.\ \eqref{14}, the angular velocities of the two colliding disks are unaffected by the collision, as expected.

\subsection{Energy dissipation}
While linear and angular momenta are conserved by collisions, kinetic energy is not. Let us see this point in more detail.
From Eqs.\   \eqref{13b&14} and  \eqref{15}, it follows that the collisional changes of $\ma\cca$, $\Ia\wa$, $\ma\ca^2$, and $\Ia\wa^2$ are
\begin{subequations}
\label{15b-Z1}
\begin{equation}
\frac{\ma \cca'-\ma\cca}{\mab}=-\enn (\gh\cdot\kk)\kk-\ett\left(\gh\cdot\kk_\perp-\Sab\right)\kk_\perp,
\label{15b}
\end{equation}
\begin{equation}
\frac{\Ia \wa'-\Ia\wa}{\mab}=\frac{\da}{2}\ett\left(\gh\cdot\kk_\perp-\Sab\right),
\label{27b}
\end{equation}
\begin{align}
\frac{\ma {\ca'}^2-\ma\ca^2}{\mab}=&\frac{\mab\enn^2}{\ma}(\gh\cdot\kk)^2-2\enn (\gh\cdot\kk)(\cca\cdot\kk)\nn
&-2\ett\left(\gh\cdot\kk_\perp-\Sab\right)(\cca\cdot\kk_\perp)
\nn
& +\frac{\mab\ett^2}{\ma}\left(\gh\cdot\kk_\perp-\Sab\right)^2,
\label{15c}
\end{align}
\begin{align}
\frac{\Ia{\wa'}^2-\Ia\wa^2}{\mab}=&
\frac{\mab\ett^2}{\ma\qa}\left(\gh\cdot\kk_\perp-\Sab\right)^2
\nn
&+\ett\da\wa\left(\gh\cdot\kk_\perp-\Sab\right).
\label{Z1}
\end{align}
\end{subequations}
Similar expressions are obtained for particle $j$ by exchanging $i\leftrightarrow j$,  $\kk\leftrightarrow -\kk$, and $\kk_\perp\leftrightarrow -\kk_\perp$.
The total kinetic energy before collision is
\begin{equation}
E_{ij}=\frac{\ma}{2}\ca^2+\frac{\mb}{2}\cb^2+\frac{\Ia}{2}\wa^2+\frac{\Ib}{2}\wb^2.
\label{Z2}
\end{equation}
Combining Eqs.\ \eqref{15c} and \eqref{Z1}, plus their counterparts for particle $j$, one obtains
\begin{align}
E_{ij}'-E_{ij}=&
-\frac{\mab}{2}\frac{\qab}{1+\qab}\left(1-\eet^2\right)\left(\gh\cdot\kk_\perp-\Sab\right)^2\nn
&-\frac{\mab}{2}\left(1-\een^2\right)(\gh\cdot\kk)^2.
\label{29}
\end{align}
The right-hand side is a negative definite quantity. Thus, energy is conserved only if the disks are elastic  ($\een=1$)  and  either perfectly smooth ($\eet=-1$) or perfectly rough ($\eet=1$). Otherwise, $E_{ij}'<E_{ij}$ and kinetic energy is dissipated upon collisions.

\subsection{Restituting collisions}
By inverting  the \emph{direct} collisional rules given by Eq.\ \eqref{13b&14} and \eqref{15}, one can find the \emph{restituting} collisional rules as
\begin{subequations}
\begin{equation}
\cca''=\cca-\frac{1}{\ma}\JJw,\quad \ccb''=\ccb+\frac{1}{\mb}\JJw,
\label{13brest}
\end{equation}
\begin{equation}
\wa''=\wa+\frac{\da}{2\Ia}\JJw\cdot\kk_\perp, \quad \wb''=\wb+\frac{\db}{2\Ib}\JJw\cdot\kk_\perp,
\label{14rest}
\end{equation}
\end{subequations}
where
\begin{equation}
\frac{\JJw}{\mab}=\frac{\enn}{\een} (\gh\cdot\kk)\kk+\frac{\ett}{\eet}\left(\gh\cdot\kk_\perp-\Sab\right)\kk_\perp.
\label{15rest}
\end{equation}
Here, the double primes denote pre-collisional quantities giving rise to unprimed quantities as post-collisional values.

It is interesting to note that the modulus of the Jacobian of the transformation between pre- and post-collisional velocities is
\begin{equation}
\left|\frac{\partial(\cca',\wa',\ccb',\wb')}{\partial(\cca,\wa,\ccb,\wb)}\right|=
\left|\frac{\partial(\cca,\wa,\ccb,\wb)}{\partial(\cca'',\wa'',\ccb'',\wb'')}\right|={\een|\eet|}.
\label{Jacob}
\end{equation}
Interestingly, this differs from the  case of spheres, for which  the Jacobian is $\een\eet^2$ \cite{SKG10}.

\section{Collisional rates of change}
\label{sec3}

\subsection{One- and two-body distribution functions}

By starting from the Liouville equation, making use of the collisional rules, and following standard steps, one can derive the Bogoliubov--Born--Green--Kirkwood--Yvon (BBGKY) hierarchy \cite{BDS97}, whose first equation reads
\begin{equation}
\partial_t \fa(\mathbf{r}_i,\xx_i;t)+\cca\cdot\nabla \fa (\mathbf{r}_i,\xx_i;t)
=\sum_{\be=1}^\NN \Qab[\mathbf{r}_i,\xx_i;t|\fab],
\label{2}
\end{equation}
where the short-hand notation $\xx_i\equiv\{\cca,\wa\}$ has been introduced,
$\fab(\mathbf{r}_i,\xx_i;\mathbf{r}_j,\xx_j;t)$ is the \emph{two-body} distribution function, and
\begin{equation}
\fa(\mathbf{r}_i,\xx_i;t)=N_\be^{-1}\int d\mathbf{r}_j\int d\xx_j\,\fab(\mathbf{r}_i,\xx_i;\mathbf{r}_j,\xx_j;t)
\label{III.1}
\end{equation}
is the \emph{one-body} distribution function, normalized as
$\int d\mathbf{r}_\al\int d \xx_i \,\fa(\mathbf{r}_i,\xx_i;t)=N_\al$. Here, $N_\al$ is the number of disks of component $\al$ and $\int d\xx_i\equiv \int d\cca\int_{-\infty}^\infty d\wa$.
Finally, the collision operator is
\begin{align}
\Qab[\mathbf{r}_i,\xx_i;t|\fab]=&\dab\int d\xx_j\int_+ d\kk\,(\gh\cdot\kk)\Bigg[\frac{1}{\een^2|\eet|}\nn
&\x\fab(\mathbf{r}_i,\xx_i'';\mathbf{r}_i-\bm{\sigma}_{ij},\xx_j'';t)\nn
&-
\fab(\mathbf{r}_i,\xx_i;\mathbf{r}_i+\bm{\sigma}_{ij},\xx_j;t)\Bigg],
\label{III.2}
\end{align}
where $\dab\equiv (\da+\db)/2$, $\bm{\sigma}_{ij}\equiv \dab\kk$, and  $\int_+ d\kk\,\equiv \int d\kk\,\Theta(\gh\cdot\kk)$, $\Theta(x)$ being the Heaviside step function.

\subsection{Balance equations}
Given a one-body  function $\psi(\xx_i)$, its average value is
\begin{equation}
\langle \psi(\xx_i)\rangle\equiv \frac{1}{\na}\int d\xx_i\, \psi(\xx_i) \fa(\xx_i),
\label{III.3}
\end{equation}
where
$
\na=\int d\xx_i\,  \fa(\xx_i)
$
is the number density of component $\al$ and, for the sake of brevity, the spatial and temporal arguments have been omitted.
In particular, one can define \emph{partial} temperatures associated with the translational and rotational degrees of freedom of each component as
\begin{equation}
\Tat=\frac{\ma}{2}\langle (\cca-\mathbf{u})^2\rangle,\quad \Tar={\Ia}\langle \wa^2\rangle,
\label{III.11}
\end{equation}
where
\begin{equation}
\mathbf{u}=\frac{\sum_{i=1}^\NN \ma\na\langle \cca\rangle}{\sum_{i=1}^\NN \ma\na}
\label{III.12}
\end{equation}
is the flow velocity. Note that in the definition of $\Tar$  the angular velocities are not referred to any average value because of the lack of invariance of the collision rules under the addition of a common value to every angular velocity. Also, Eq.\ \eqref{III.11} takes into account that the number of translational and rotational degrees of freedom are $2$ and $1$, respectively.
The \emph{global} temperature is
\begin{equation}
T=\sum_{\al=1}^\NN\frac{\na}{n}\frac{2\Tat+\Tar}{3},
\label{III.13}
\end{equation}
where $n=\sum_{\al=1}^\NN \na$ is the total number density.

In general, the balance equation for $\langle \psi(\xx_i)\rangle$ can be obtained by multiplying both sides of Eq.\ \eqref{2} by $\psi(\xx_i)$ and integrating over $\xx_i$:
\begin{equation}
\partial_t \na \langle \psi(\xx_i)\rangle+\nabla\cdot \na \langle \cca\psi(\xx_i)\rangle=\sum_{\be=1}^\NN \Iab[\psi|\fab],
\label{III.5}
\end{equation}
where the collisional integral $\Iab[\psi|\fab]$ is
\begin{align}
\Iab[\psi|\fab]\equiv&\int d\xx_i\, \psi(\xx_i) \Qab[\xx_i|\fab]\nn
=&\dab\int d\xx_i \int d\xx_j\int_+ d\kk\, (\gh\cdot\kk)
\nn
&\x
\fab(\mathbf{r}_\al,\xx_i;\mathbf{r}_i+\bm{\sigma}_{ij},\xx_j)
\left[\psi(\xx_i')-\psi(\xx_i)\right].
\label{3}
\end{align}
Therefore, $\na^{-1}\Iab[\psi|\fab]$ is the \emph{rate of change} of the quantity $\psi(\xx_i)$ due to collisions with particles of component $\be$. This rate of change is a functional of the two-body distribution function $\fab$, as indicated by the notation.
The most basic cases are $\psi(\xx_i)=\{\ma \cca,\Ia \wa, \ma \ca^2,\Ia \wa^2\}$. The corresponding rates of change are obtained by inserting Eqs.\  \eqref{15b-Z1} into Eq.\ \eqref{3}. Note that so far all the results are formally exact.

\subsection{Collisional integrals as two-body averages}
To proceed, let us make the approximation
\begin{equation}
\Iab[\psi|\fab]\approx  \Iab[\psi|\ffab],
\label{III.6}
\end{equation}
where
\begin{align}
\label{III.7}
\ffab(\mathbf{r}_\al,\xx_i;\xx_j)\equiv&
\frac{1}{\int_+ d\kk\, (\gh\cdot\kk)}\int_+ d\kk\, (\gh\cdot\kk)\nn
&\x\fab(\mathbf{r}_\al,\xx_i;\mathbf{r}_i+\bm{\sigma}_{ij},\xx_j),
\end{align}
is the orientational average  of the \emph{pre-collisional} distribution $\fab$.
Equation \eqref{III.6} replaces the formally exact collisional integral \eqref{3} by a simpler one where the angular integral
\begin{equation}
\Psi(\xx_i;\xx_j)\equiv \int_+ d\kk\, (\gh\cdot\kk)
\left[\psi(\xx_i')-\psi(\xx_i)\right]
\label{III.8}
\end{equation}
can be evaluated independently of $\fab$. As a consequence,
\begin{equation}
\Iab[\psi|\ffab]=\na\nb\dab\llangle \Psi(\xx_i;\xx_j)\rrangle,
\end{equation}
where
\begin{equation}
\llangle \Psi(\xx_i;\xx_j)\rrangle\equiv \frac{1}{\na \nb}\int d\xx_i \int d\xx_j\,\Psi(\xx_i;\xx_j)\ffab(\xx_i;\xx_j)
\label{17}
\end{equation}
is a \emph{two-body} average.

It is important to bear in mind  that the approximation \eqref{III.6} refers to pre-collisional quantities  inside integrals over $\xx_i$, $\xx_j$, and $\kk$. Thus, it is much weaker than the bare approximation $\fab\approx\ffab$.
On the other hand, it must be pointed out that the equality $\fab=\ffab$ holds if
(i) the gas is in the Boltzmann limit ($\na\da^2\to 0$, $\nb\db^2\to 0$), in which case one can formally take $\dab\to 0$ in the contact value of $\fab$, or (ii) the system is homogeneous and isotropic (regardless of the reduced densities $\na\da^2$ and $\nb\db^2$), in which case $\fab$ only depends on $|\mathbf{r}_\al-\mathbf{r}_\be|$.
Thus, the approximation \eqref{III.6} is justified if the density of the granular gas and/or its heterogeneities are small enough to make the value of $\fab$ at contact hardly dependent on the relative orientation of the two colliding disks.

\begin{table}[tbp]
\caption{Relevant collisional integrals  in terms of two-body averages.}
\label{table:0}
\begin{ruledtabular}
\begin{tabular}{ll}
$\psi$ & $-\Iab[\psi|\ffab]/\mab\na\nb\dab$\\
\hline
\vspace{-4mm}
\\
$\ma\cca$&$\displaystyle{\frac{2}{3}\left(2\enn+\ett\right)\llangle \g\gh\rrangle-\frac{\pi}{2}\ett\llangle \Sab{\gh}_\perp\rrangle}$\\
\vspace{-4mm}
\\
$\Ia\wa$&$\displaystyle{\da\ett\llangle \g\Sab\rrangle}$\\
\vspace{-4mm}
\\
$\ma\ca^2$&$\displaystyle{\frac{4}{3}\left(2\enn+\ett\right)\llangle \g\cca\cdot\gh\rrangle
+{\pi} \ett \llangle \Sab   \cca\cdot {\gh}_\perp\rrangle}$\\
&$\displaystyle{-
\frac{2\mab}{3\ma}\left(2\enn^2+ \ett^2\right)\llangle\g^3\rrangle
-\frac{2\mab \ett^2}{\ma}\llangle\g\Sab^2\rrangle}$\\
\vspace{-4mm}
\\
$\Ia\wa^2$&$\displaystyle{2\ett\da\llangle \g\wa\Sab\rrangle -\frac{2\mab \ett^2}{3\ma\qa}\left(\llangle
\g^3\rrangle+3\llangle\g\Sab^2\rrangle\right)}$\\
\vspace{-4mm}
\\
$\ma\ca^2+\mb\cb^2$&$\displaystyle{\frac{4}{3}\left[(1-\een^2)+\frac{\ett}{2}(2-\ett)\right]\llangle \g^3\rrangle
-{2\ett^2}\llangle\g\Sab^2\rrangle}$\\
\vspace{-4mm}
\\
$\Ia\wa^2+\Ib\wb^2$&$\displaystyle{\frac{2\ett}{\qab}\left(2\qab-{\ett}\right)\llangle\g\Sab^2\rrangle
-\frac{2}{3\qab}\ett^2\llangle
\g^3\rrangle}$\\
\vspace{-4mm}
\\
$E_{ij}$&$\displaystyle{\frac{2}{3}(1-\een^2)\llangle \g^3\rrangle+\frac{\qab}{3(1+\qab)}(1-\eet^2)
}$\\
&$\displaystyle{\times\left({\llangle \g^3\rrangle}+3\llangle\g\Sab^2\rrangle\right)}$\\
\end{tabular}
\end{ruledtabular}
\end{table}

Let us now particularize to $\psi(\xx_i)=\{\ma\cca, \Ia\wa, \ma\ca^2, \Ia\wa^2\}$. The needed angular integrals are
\begin{subequations}
\label{6--10}
\begin{equation}
\int_+ d\kk \,(\uk\cdot\kk)^\ell\kk=\frac{\sqrt{\pi}\Gamma(1+\ell/2)}{\Gamma\left(\frac{\ell+3}{2}\right)} \uk,
\label{7}
\end{equation}
\begin{equation}
\int_+ d\kk\, (\uk\cdot\kk)\kk_\perp=\frac{\pi}{2}\uk_\perp,
\label{9}
\end{equation}
\begin{equation}
\int_+ d\kk\, (\uk\cdot\kk)(\uk\cdot\kk_\perp)^\ell\kk_\perp=\frac{1+(-1)^{\ell+1}}{\ell+2}\uk,
\label{10}
\end{equation}
\end{subequations}
were $\uk$ is an arbitrary unit vector and  $\uk_\perp=\uk\x\widehat{\mathbf{z}}$ is its orthogonal unit vector.
After some algebra, one can find the expressions displayed in  Table \ref{table:0},
where ${\gh}_\perp=\gh\x\widehat{\mathbf{z}}$ is a vector orthogonal to $\gh$.

\subsection{Estimates of two-body averages}
\label{sec4}
Table \ref{table:0} expresses the collisional rates of change of the main quantities as linear combinations of two-body averages of the form \eqref{17}. They are local functions of space and time and functionals of the orientation-averaged  pre-collisional distribution $\ffab$. While, thanks to the approximation \eqref{III.6}, the expressions in Table \ref{table:0} are much more explicit than the formally exact results stemming from Eq.\ \eqref{3}, they still require the full knowledge of  $\ffab$.

\begin{table}[tbp]
\caption{Expressions, as obtained from the approximation \protect\eqref{IV.1}, for the two-body averages appearing in Table \ref{table:0}.}
\label{table:1}
\begin{ruledtabular}
\begin{tabular}{rl}
Quantity & Expression \\ \hline
$\llangle \g\gh\rrangle$&$\mathbf{0}$\\
$\llangle \Sab{\gh}_\perp\rrangle$&$\mathbf{0}$\\
$\llangle \g\Sab\rrangle$&$ \displaystyle{\frac{1}{2}\left(\da {\Omega}_\al+\db {\Omega}_\be\right)}\llangle \g\rrangle$\\
$\llangle \g\cca\cdot\gh\rrangle$&$\displaystyle{\frac{\Tat}{\ma}\left(\frac{\Tat}{\ma}+\frac{\Tbt}{\mb}\right)^{-1}}
\llangle\g^3\rrangle$\\
$\llangle \Sab   \cca\cdot {\gh}_\perp\rrangle$&$0$\\
$\llangle\g\Sab^2\rrangle$&$\displaystyle{\left(\frac{\Tar}{\ma\qa}+\frac{\Tbr}{\mb\qb}+\frac{1}{2}\da\db {\Omega}_\al {\Omega}_\be\right)}\llangle\g\rrangle$\\
$\llangle \g\wa\Sab\rrangle$&$\displaystyle{\left(\frac{2\Tar}{\ma\qa\da}+\frac{1}{2}\db {\Omega}_\al {\Omega}_\be\right)}\llangle \g\rrangle$\\
$\llangle\g\rrangle$&$\displaystyle{\sqrt{\frac{\pi}{2}}\chiab\left(\frac{\Tat}{\ma}+\frac{\Tbt}{\mb}\right)^{1/2}}$\\
$\llangle\g^3\rrangle$&$\displaystyle{3\sqrt{\frac{\pi}{2}}\chiab\left(\frac{\Tat}{\ma}+\frac{\Tbt}{\mb}\right)^{3/2}}$\\
\end{tabular}
\end{ruledtabular}
\end{table}

Suppose, for simplicity, that $\langle \cca\rangle=\langle \ccb\rangle=\mathbf{u}$.
Now, let us imagine that, instead of the full knowledge of $\ffab$, we only know the  common flow velocity ($\mathbf{u}$) and the two translational temperatures ($\Tat$ and $\Tbt$). One can resort to information-theory (i.e., maximum-entropy) arguments to make the approximation
\begin{align}
\ffab(\xx_i;\xx_j)\to&  \frac{\chiab\ma\mb}{4\pi^2\Tat\Tbt}e^{-{\ma(\cca-\mathbf{u})^2}/{2\Tat}}\far(\wa)\nn
&\x e^{-{\mb(\ccb-\mathbf{u})^2}/{2\Tbt}}
\fbr(\wb),
\label{IV.1}
\end{align}
where $\chiab$ is the contact value of the pair correlation function and
\begin{equation}
\far(\wa)=\int d\cca\,\fa(\xx_i)
\end{equation}
is the marginal distribution function associated with the rotational degrees of freedom. Similarly, the translational marginal distribution function is
\begin{equation}
\fat(\cca)=\int_{-\infty}^\infty d\wa\,\fa(\xx_i).
\end{equation}
Equation \eqref{IV.1}  is the least biased ansatz consistent with the input quantities  $\mathbf{u}$, $\Tat$, and $\Tbt$. It implies (a) molecular chaos (i.e., $\ffab=\chiab \fa\fb$),  (b) statistical independence between the translational and angular velocities (i.e., $\fa=\na^{-1}\fat\far$), and (c) a Maxwellian form for the distribution of translational velocities.
The generalization to $\langle \cca\rangle\neq\langle \ccb\rangle$ can be carried out following similar steps as  done in Ref.\  \cite{VGS07} for smooth spheres.
Since the angular velocities only appear linearly or quadratically in  Table \ref{table:0}, a Maxwellian form for $\far$ does not need to be assumed, so that the local densities ($\na$ and $\nb$), the average angular velocities ($\langle \wa\rangle={\Omega}_\al$ and $\langle \wb\rangle={\Omega}_\be$),
and the rotational temperatures ($\Tar$ and  $\Tbr$) do not appear explicitly in Eq.\ \eqref{IV.1}.

It must be stressed that, while small deviations from the three assumptions  (a), (b), and (c) behind Eq.\ \eqref{IV.1} have been documented in the literature \cite{SM01,SPM01,BPKZ07,SKS11,VSK14}, the expectation is that  the two-body averages can be estimated reasonably well by performing the replacement \eqref{IV.1}. This expectation has been confirmed in the hard-sphere case \cite{VSK14,VLSG17,VLSG17b}.

The insertion of the approximation \eqref{IV.1} into Eq.\ \eqref{17} for the functions $\Psi(\xx_i;\xx_j)$ appearing in Table \ref{table:0} yields the results displayed in Table \ref{table:1}. In particular, combining the second row of Table \ref{table:0} with the third and eighth rows of Table \ref{table:1}, it is straightforward to obtain
\begin{equation}
\Iab[\Ia\wa]=-\frac{1}{4}\na{\nu_{\al\be}}\mab{\ett}\da\left(\da{\Omega}_\al+\db{\Omega}_\be\right),
\label{V.1}
\end{equation}
where  the effective collision frequency
\begin{equation}
\nu_{\al\be}\equiv\sqrt{2\pi}\chiab\nb{\dab}\sqrt{\frac{\Tat}{\ma}+\frac{\Tbt}{\mb}}
\label{56b}
\end{equation}
has been introduced. Equation \eqref{V.1} shows that, except in the smooth case ($\beta_{ij}=-1$), collisions produce a systematic decrease in the magnitude of the angular velocities of the particles. In the monodisperse case, the collision frequency \eqref{56b} reduces to
\begin{equation}
\label{numono}
\nu=2\chi n\sigma\sqrt{\pi T^\tr/m}.
\end{equation}

\begin{table*}[tbp]
\caption{Energy production rates ($\xi$s), cooling rates ($\zeta$s), and equipartition rates ($\Xi$s) for polydisperse and monodisperse systems.}
\label{table:2}
\begin{ruledtabular}
\begin{tabular}{ll}
\multicolumn{2}{c}{Polydisperse system}\\
Quantity & Expression\\
\hline
\vspace{-4mm}
\\
$\zabt$&$\displaystyle{\frac{\nu_{\al\be}\mab^2}{\ma\Tat}
\left[(2\enn+\ett)\frac{\Tat}{\mab}-
\frac{2\enn^2+ \ett^2}{2}\left(\frac{\Tat}{\ma}+\frac{\Tbt}{\mb}\right)
-\frac{\ett^2}{2}\left(\frac{\Tar}{\ma\qa}+\frac{\Tbr}{\mb\qb}+\frac{1}{2}\da\db {\Omega}_\al{\Omega}_\be\right)
\right]}$\\
\vspace{-4mm}
\\
$\zabr$&$\displaystyle{\frac{\nu_{\al\be}\mab^2 \ett}{\ma\qa\Tar}\left[\frac{2\Tar}{\mab}+\frac{\ma}{2\mab}\qa\da\db {\Omega}_\al{\Omega}_\be-{ \ett}\left(\frac{\Tat}{\ma}+\frac{\Tbt}{\mb}+\frac{\Tar}{\ma\qa}+\frac{\Tbr}{\mb\qb}+\frac{1}{2}\da\db {\Omega}_\al{\Omega}_\be\right)\right]}$\\
\vspace{-4mm}
\\
$\zeta$&$\displaystyle{\sum_{\al,\be=1}^\NN \frac{\na\nu_{\al\be}\mab}{3nT}\left[(1-\een^2)\left(\frac{\Tat}{\ma}+\frac{\Tbt}{\mb}\right)
+\frac{\qab(1-\eet^2)}{2(1+\qab)}
\left(\frac{\Tat}{\ma}+\frac{\Tbt}{\mb}+\frac{\Tar}{\ma\qa}+\frac{\Tbr}{\mb\qb}+\frac{1}{2}\da\db {\Omega}_\al {\Omega}_\be\right)\right]}$\\
\vspace{-4mm}
\\
$\zeta_{\al\be}^\tr$&$\displaystyle{\frac{\nu_{ij}\mab^2(1-\een^2)}{\ma\Tat}\left(\frac{\Tat}{\ma}+\frac{\Tbt}{\mb}\right)}$\\
$\zeta_{\al\be}^\rot$&$\displaystyle{\frac{\nu_{ij}\mab^2\qab^2(1-\eet^2)}{\ma\qa(1+\qab)^2\Tar}\left(\frac{\Tat}{\ma}+\frac{\Tbt}{\mb}+
\frac{\Tar}{\ma\qa}+\frac{\Tbr}{\mb\qb}+\frac{1}{2}\da\db\Omega_i\Omega_j\right)}$\\
\vspace{-4mm}
\\
$\Xi_{ij}^{(1)}$&$\displaystyle{\frac{2\nu_{ij}\mab^2(1+\een)}{\ma\mb\Tat}\left(\Tat-\Tbt\right)}$\\
\vspace{-4mm}
\\
$\Xi_{ij}^{(2)}$&$\displaystyle{\frac{\nu_{ij}\mab\qab(1+\eet)}{\ma(1+\qab)\Tat}\left(
\Tat-\Tar-\frac{\ma\qa\da\db\Omega_i\Omega_j}{4}\right)}$\\
\vspace{-4mm}
\\
$\Xi_{ij}^{(3)}$&$\displaystyle{\frac{2\nu_{ij}\mab^2\qab^2(1+\eet)}{(1+\qab)^2\Tar}\left[\frac{\Tar-\Tbr}{\ma\mb\qa\qb}+\frac{\Tat-\Tbt}{\ma\mb\qa}
+\frac{\Tar-\Tat}{\ma\mab\qa}+\left(\frac{\ma\qa-\mb\qb}{\ma\mb\qa\qb}+\frac{1}{\mab}\right)\frac{\da\db\Omega_i\Omega_j}{4}
\right]}$\\
\\
\hline
\multicolumn{2}{c}{Monodisperse system}\\
Quantity & Expression\\
\hline
\vspace{-4mm}
\\
$\xi^\tr$&$\displaystyle{\frac{{\nu}\kappa}{(1+\kappa)^2}\frac{1+\beta}{2}\left[1-\frac{T^\rot+{\kappa}m\sigma^2\Omega^2/4}{T^\tr}
+\frac{1-\beta}{2}\left(\kappa+\frac{T^\rot+{\kappa}m\sigma^2\Omega^2/4}{T^\tr}\right)\right]+\nu\frac{1-\alpha^2}{2}}$\\
\vspace{-4mm}
\\
$\xi^\rot$&$\displaystyle{\frac{{\nu}\kappa}{(1+\kappa)^2}(1+\beta)\frac{T^\tr}{T^\rot}\left[\frac{T^\rot+{\kappa}m\sigma^2\Omega^2/4}{T^\tr}-1+\frac{1-\beta}{2\kappa}\left(\kappa+\frac{T^\rot+{\kappa}m\sigma^2\Omega^2/4}{T^\tr}\right)\right]}$\\
\vspace{-4mm}
\\
$\zeta$&$\displaystyle{\frac{\nu T^\tr}{2T^\tr+T^\rot}\left[ \frac{1-\beta^2}{2(1+\kappa)}\left(\kappa+\frac{T^\rot+{\kappa}m\sigma^2\Omega^2/4}{T^\tr}\right)+1-\alpha^2\right]}$\\
\vspace{-4mm}
\\
$\zeta^\tr$&$\displaystyle{\nu\frac{1-\alpha^2}{2}}$\\
\vspace{-4mm}
\\
$\zeta^\rot$&$\displaystyle{\frac{\nu}{(1+\kappa)^2}\frac{1-\beta^2}{2}\frac{T^\tr}{T^\rot}\left(\kappa+\frac{T^\rot+{\kappa}m\sigma^2\Omega^2/4}{T^\tr}\right)}$\\
$\Xi^{(1)}$&$0$\\
\vspace{-4mm}
\\
$\Xi^{(2)}$&$\displaystyle{\frac{\nu\kappa}{1+\kappa}\frac{1+\beta}{2}\left(1-\frac{T^\rot+{\kappa}m\sigma^2\Omega^2/4}{T^\tr}\right)}$\\
$\Xi^{(3)}$&$\displaystyle{-\frac{2}{1+\kappa}\frac{T^\tr}{T^\rot}\Xi^{(2)}}$\\
\end{tabular}
\end{ruledtabular}
\end{table*}

\section{Energy production  rates and cooling rate}
\label{sec5}
While part of the total kinetic energy is dissipated after each collision [see Eq.\ \eqref{29}],  each one of the four partial kinetic energy contributions in Eq.\ \eqref{Z2} can either increase or decrease after a given collision, as a consequence of a redistribution of the non-dissipated energy among both colliding particles and both types (translational and rotational) of energy. To characterize the statistical effect of energy dissipation and redistribution, let us introduce the energy production rates as the rates of change of the partial temperatures $\Tat$ and $\Tar$ due to collisions of disks of component $\al$ with disks of component $\be$:
\begin{equation}
\zabt\equiv -\frac{\Iab[\ma(\cca-\mathbf{u})^2]}{2\na\Tat},
\quad \zabr\equiv -\frac{\Iab[\Ia\wa^2]}{\na\Tar}.
\end{equation}
When collisions of particles of component $\al$ with all the components  are considered, one gets the (total) energy production rates
\begin{subequations}
\label{107}
\begin{equation}
\xi^\tr_\al\equiv-\frac{1}{\Tat}\left(\frac{\partial\Tat}{\partial t}\right)_{\text{coll}}=\sum_{\be=1}^\NN \zabt,
\end{equation}
\begin{equation}
\xi^\rot_\al\equiv-\frac{1}{\Tar}\left(\frac{\partial\Tar}{\partial t}\right)_{\text{coll}}=\sum_{\be=1}^\NN \zabr.
\end{equation}
\end{subequations}
Finally, the net \emph{cooling} rate is
\begin{equation}
\zeta\equiv-\frac{1}{T}\left(\frac{\partial T}{\partial t}\right)_{\text{coll}}=\sum_{\al=1}^\NN \frac{\na}{n}
\frac{2\Tat\xi^\tr_\al+\Tar\xi^\rot_\al}{3T}.
\label{110}
\end{equation}
As said before, the individual energy productions rates $\zabt$ and $\zabr$ (or even $\xi^\tr_\al$ and $\xi^\rot_\al$) do not have a definite sign. In contrast, the net cooling rate $\zeta$ must be positive definite, i.e., collisions produce a decrease of the total temperature $T$ unless $\een=1$ and $\eet=\pm 1$ for \emph{all} pairs $\al\be$.

{}The combination of the expressions in Tables \ref{table:0} and \ref{table:1} allows one to obtain the energy production rates $\zabt$ and $\zabr$, and the cooling rate $\zeta$. The resulting expressions can be seen in the first half of Table \ref{table:2} as explicit functions of the local values of  $\na$, $\nb$, ${\Omega}_\al$, ${\Omega}_\be$, $\Tat$, $\Tar$, $\Tbt$, and $\Tbr$, as well as of the mechanical parameters $\ma$, $\mb$, $\da$, $\db$, $\qa$, $\qb$, $\een$, and $\eet$.

In the expressions for $\zabt$ and $\zabr$ given in Table \ref{table:2}, the dissipation and redistribution effects are mixed together. To disentangle them, it is convenient to carry out the decompositions \cite{S11b}
\begin{subequations}
\begin{equation}
\label{B4}
\zabt=\frac{\qa\Tar}{2\Tat}\zabr+\zeta_{\al\be}^\tr+\Xi_{ij}^{(1)}
+\Xi_{ij}^{(2)},
\end{equation}
\begin{equation}
\label{B1}
\zabr=\zeta_{\al\be}^\rot+\Xi_{ij}^{(3)},
\end{equation}
\end{subequations}
where the expressions for $\zeta_{\al\be}^\tr$, $\zeta_{\al\be}^\rot$, and $\Xi_{ij}^{(1\text{--}3)}$ are also included in Table \ref{table:2}.

The quantities $\Xi_{\al\be}^{(1\text{--}3)}$ represent \emph{equipartition} rates. They do not have a definite sign and vanish if all the temperatures are equal and either $\Omega_i=0$ or $\Omega_j=0$. The equipartition rate $\Xi_{ij}^{(1)}$ is always present (even for perfectly elastic disks, $\een=1$) and tends to equilibrate the translational temperatures $\Tat$ and $\Tbt$. The rates $\Xi_{\al\be}^{(2)}$ and $\Xi_{\al\be}^{(3)}$ do not contribute in the case of smooth spheres ($\eet=-1$). The former tends to equilibrate the translational ($\Tat$) and rotational ($\Tar$) temperatures of component $\al$, while the latter tends to equilibrate the rotational temperatures $\Tar$ and $\Tbr$ but is also affected by the other temperature differences ($\Tat-\Tbt$ and $\Tat-\Tar$), and by the product $\Omega_i\Omega_j$. On the other hand, the quantities $\zeta_{\al\be}^{\tr}$ and $\zeta_{\al\be}^{\rot}$ are positive definite and represent \emph{cooling} rates. The former (headed by ${1-\een^2}$) vanishes only if the spheres are elastic, while the latter (headed by ${1-\eet^2}$) vanishes only if the spheres are  either perfectly smooth ($\eet=-1$) or perfectly rough ($\eet=1$).

It is straightforward to check that ${\na}\Tat\Xi_{\al\be}^{(1)}+{\nb}\Tbt\Xi_{\be\al}^{(1)}=0$ and $\na\left[2\Tat\Xi_{\al\be}^{(2)}+(1+\qa)\Tar\Xi_{\al\be}^{(3)}\right]
+\nb\left[2\Tbt\Xi_{\be\al}^{(2)}+(1+\qb)\Tbr\Xi_{\be\al}^{(3)}\right]=0$.
Therefore, as expected, the equipartition rates $\Xi_{ij}^{(1\text{--}3)}$ do not contribute to the net cooling rate $\zeta$ defined by Eq.\ \eqref{110}, so that
\begin{align}
\label{B8}
\zeta=&\frac{1}{3nT}\sum_{i,j=1}^\NN\left[\na\left(\Tat\zeta_{ij}^\tr+\frac{1+\qa}{2}\Tar\zeta_{ij}^\rot\right)\right.\nn
&\left.+\nb\left(\Tbt\zeta_{ji}^\tr+\frac{1+\qb}{2}\Tbr\zeta_{ji}^\rot\right)\right].
\end{align}

In the monodisperse limit (i.e., $s=1$ or, equivalently, $m_i=m$, $\qa=\kappa$, $\da=\sigma$, $\een=\alpha$, $\eet=\beta$, $\Tat=T^\tr$, $\Tar=T^\rot$, $\Omega_i=\Omega$, $\na=n$, $\chi_{ij}=\chi$), the energy production, cooling, and equipartition rates simplify to the expressions shown in the second half of Table \ref{table:2}, in agreement with previous results \cite{LHMZ98}. Moreover, particularization of the expressions presented in Table \ref{table:2} to the case of multicomponent smooth disks ($\eet=-1$) allows one to recover known results  \cite{VGS07}.

The expressions displayed in Table \ref{table:2} are the main results of this paper. As an immediate application, the HCS is analyzed in Secs.\ \ref{sec6} and \ref{sec7}.

\section{Application to the homogeneous cooling state}
\label{sec6}
The HCS is an isotropic and spatially uniform freely cooling regime, reached after the influence of the initial preparation has vanished. This base state has been experimentally realized in conditions of microgravity or levitation \cite{MIMA08,GBG09,TMHS09,HTMWS15,HTWS18}. As a consequence of isotropy, the mean angular velocities are zero (i.e., $\Omega_i=0$), while, as a consequence of spatial uniformity, the flux term $\nabla\cdot \na \langle \cca\psi(\cca,\wwa)\rangle$ in Eq.\ \eqref{III.5} is absent. Therefore, the evolution equations for the total and partial temperatures are
\begin{subequations}
\begin{equation}
 \partial_t T=-\zeta T,
\label{57.1}
\end{equation}
\begin{equation}
 \partial_t\frac{\Tat}{T}=-\left(\zt_\al-\zeta\right)\frac{\Tat}{T},
 \quad \partial_t\frac{\Tar}{T}=-\left(\zr_\al-\zeta\right)\frac{\Tar}{T}.
\label{57.2}
\end{equation}
\end{subequations}

Once the HCS scaling regime is reached (after a certain transient time), all the time dependence of the gas occurs through the total temperature $T$. This implies constant temperature ratios and equal production rates, i.e.,
\begin{subequations}
\label{59}
\begin{equation}
\zt_1=\zt_2=\cdots=\zt_\NN,
\quad \zr_1=\zr_2=\cdots=\zr_\NN,
\label{59a}
\end{equation}
\begin{equation}
\zt_1=\zr_1.
\label{59c}
\end{equation}
\end{subequations}
When Eqs.\ \eqref{107}, together with the expressions in Table \ref{table:2}, are used in Eqs.\ \eqref{59}, the latter make a set of $2\NN-1$ equations whose solution gives the $2\NN-1$ temperature ratios  $\Tr_1/\Tt_1$ and $\{\Tt_i/\Tt_1,\Tr_i/\Tr_1;i=2,\ldots,\NN\}$  for arbitrary values of the $\NN^2+5\NN-2$ free dimensionless parameters of the problem: the total packing fraction $\phi=\frac{\pi}{4}\sum_{i=1}^sn_i\sigma_i^2$, the $\NN-1$ density ratios $\{n_i/n_1\}$, the $\NN-1$ size ratios $\{\sigma_i/\sigma_1\}$, the $\NN-1$ mass ratios $\{m_i/m_1\}$, the $\NN$ reduced moments of inertia  $\{\qa\}$, the $\NN(\NN+1)/2$ coefficients of normal restitution $\{\alpha_{ij}\}$, and the $\NN(\NN+1)/2$ coefficients of tangential restitution $\{\beta_{ij}\}$.

\subsection{Monodisperse system}
\label{subsec:Mono}

\begin{figure}
\includegraphics[width=\ancho\columnwidth]{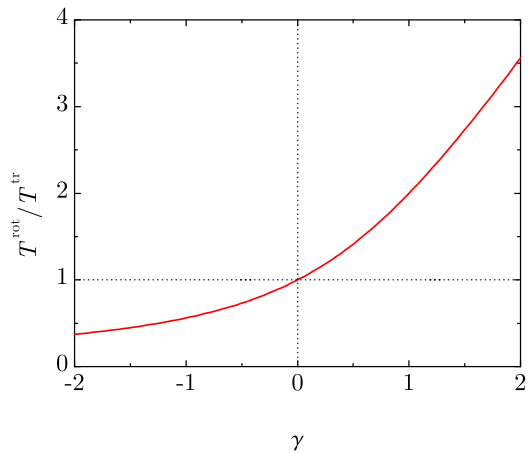}
\caption{Plot of the temperature ratio $T^\rot/T^\tr$ of the monodisperse gas versus the parameter $\gamma$ defined in Eq.\ \eqref{gamma}, according to the theoretical prediction  \eqref{Tr2}. }
\label{fig:T_mono}
\end{figure}

\begin{figure}
\includegraphics[width=\ancho\columnwidth]{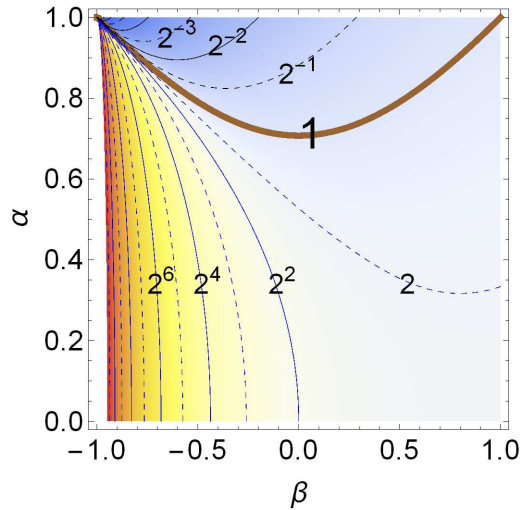}
\caption{Density plot of  $\Tr/\Tt$ [see Eqs.\ \eqref{Tr2} and \eqref{gamma}] for $\q=\frac{1}{2}$. The contour lines correspond to $\Tr/\Tt=1$ (thick solid line), $\Tr/\Tt=2^{-1}, 2^{-2}, 2^{-3}, \ldots$, and $\Tr/\Tt=2, 2^{2}, 2^{3}, \ldots$. The temperature ratio $T^\rot/T^\tr$ takes the same value for all the pairs $(\alpha,\beta)$ lying on the same locus $\gamma=\text{const}$.  }
\label{fig:Contour}
\end{figure}

In the monodisperse case ($s=1$) the only unknown is $\Tr/\Tt$ and the true number of free parameters is $3$ because the packing fraction $\phi$ is absorbed via the pair correlation function at contact, $\chi$, into the collision frequency $\nu$ [cf.\ Eq.\ \eqref{numono}]. The HCS condition $\zt=\zr$ yields a quadratic equation whose physical solution is
\begin{equation}
\label{Tr2}
\frac{\Tr}{\Tt}=\sqrt{2+\left(\gamma-\frac{1}{2}\right)^2}+\gamma-\frac{1}{2},
\end{equation}
where the parameter
\begin{equation}
\label{gamma}
\gamma\equiv \frac{(1+\kappa)^2}{\kappa(1+\beta)^2}\left[1-\alpha^2-\frac{2-\q}{2(1+\kappa)}(1-\beta^2)\right]
\end{equation}
comprises completely the dependence of the temperature ratio on the three quantities  $\alpha$, $\beta$, and $\q$. The dependence of $\Tr/\Tt$ on $\gamma$ is shown in Fig.\ \ref{fig:T_mono}.

It can be observed from Eq.\ \eqref{gamma} that the sign of $\gamma$ results from the competition between two terms: $1-\alpha^2$, on the one hand, and a term proportional to $1-\beta^2$, on the other hand. {}From Eq.\ \eqref{restitution}, it turns out that  that $1-\alpha^2=1-(\mathbf{w}'\cdot\kk )^2/(\mathbf{w}\cdot\kk)^2$ measures the relative decrease in the magnitude of the normal component of the relative velocity after a collision. Likewise,  $1-\beta^2=1-(\mathbf{w}'\cdot\kk_\perp)^2/(\mathbf{w}\cdot\kk_\perp)^2$ measures a similar relative decrease but in the case of the tangential component. Thus, $\gamma>0$ if the relative decrease of the normal component is larger than that of the tangential component (the latter being multiplied by a $\q$-dependent factor). In such a case,  ${\Tr}/{\Tt}>1$. Otherwise, if the relative decrease of the normal component is smaller than that of the ($\q$-weighted) tangential component, then $\gamma<0$ and  ${\Tr}/{\Tt}<1$. Equipartition of energy (${\Tr}/{\Tt}=1$) occurs if $\gamma=0$, implying a balance (in the sense described above) between the relative decrease of the magnitudes of the tangential and normal components of the relative velocity.
A similar dependence of $\Tr/\Tt$ on a certain single parameter $\gamma$ occurs in the case of spheres \cite{VLSG17}. A detailed comparison shows that the breakdown of rotational-translational equipartition is typically higher in disks than in spheres.

To have a more comprehensive view on the joint dependence of ${\Tr}/{\Tt}$ on the coefficients of restitution $\alpha$ and $\beta$, Fig.\ \ref{fig:Contour} shows a density plot of the temperature ratio in the case of uniform disks ($\q=\frac{1}{2}$). The equipartition line ${\Tr}/{\Tt}=1$, where $\gamma=0$ (i.e., $\alpha=\sqrt{(1+\beta^2)/2}$, with a minimum at $\alpha=1/\sqrt{2}\simeq 0.707$), splits the plane $(\beta,\alpha)$ into two regions. In the upper region ($\gamma<0$) one has ${\Tr}/{\Tt}<1$, whereas ${\Tr}/{\Tt}>1$ in the lower region ($\gamma>0$). Moreover it can be observed that ${\Tr}/{\Tt}$ grows very rapidly in the lower region as one approaches the quasismooth limit $\est \to -1$. In contrast,  ${\Tr}/{\Tt}\to 0$ in the same limit $\est \to -1$ if $\alpha=1$ (elastic collisions).
In fact, Eq.\ \eqref{gamma} yields
\begin{equation}
\lim_{\est\to -1}\gamma=
\begin{cases}
\displaystyle{\frac{(1+\q)^2}{\q}}\frac{1-\esn^2}{(1+\est)^{2}}\to \infty,&\esn<1,\\
-\displaystyle{\frac{2+\q(1-\q)}{\q(1+\est)}}\to -\infty,&\esn=1,
\end{cases}
\label{80a}
\end{equation}
so that
\begin{equation}
\lim_{\est\to -1}\frac{\Tr}{\Tt}=
\begin{cases}2\gamma=
\displaystyle{\frac{2(1+\q)^2}{\q}}\frac{1-\esn^2}{(1+\est)^2}\to \infty,&\esn<1,\\
-\gamma^{-1}=\displaystyle{\frac{\q (1+\est)}{2+\q(1-\q)}}\to 0,&\esn=1.
\end{cases}
\label{80}
\end{equation}

Therefore, the elastic-disk limit ($\esn\to 1$) and the smooth-disk limit ($\est\to -1$) do not commute. If the disks are inelastic ($\esn<1$)  and quasismooth ($\est\to -1$), the rotational and translational degrees of freedom tend to be decoupled  and $\Tr$ does not change with time, while $\Tt$ keeps decreasing due to inelasticity \cite{HHZ00}. As a consequence, the ratio $\Tr/\Tt$ diverges in the long-time limit. On the other hand, if the disks are perfectly elastic ($\esn=1$) and then the quasismooth limit ($\est\to -1$) is taken,  a nonzero coupling between $\Tr$ and $\Tt$ exists such that, assuming an initial state with $\Tr\sim\Tt$, the translational temperature decays initially more slowly than the rotational temperature and  $\Tr/\Tt$ decreases in time until the HCS condition $\zr/\zt\approx 2/\q-(\Tt/\Tr)(1+\beta)/(1+\q)=1$ eventually results in a temperature ratio $\Tr/\Tt\sim 1+\est\to 0$.

\subsection{Bidisperse system}

\begin{figure}
\includegraphics[width=\ancho\columnwidth]{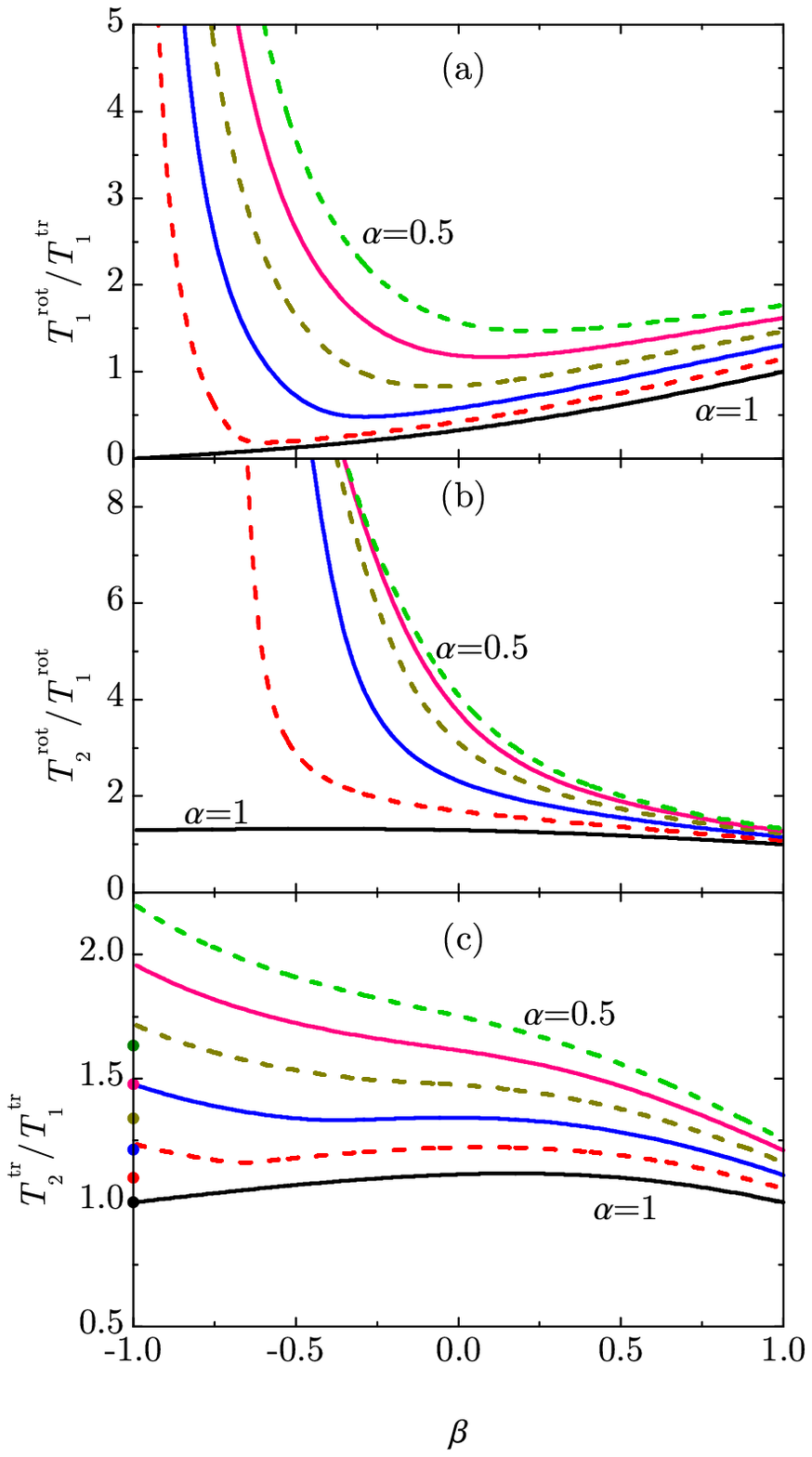}
\caption{
Plot of the temperature ratios (a) $\Tr_1/\Tt_1$, (b) $\Tr_2/\Tr_1$, and (c) $\Tt_2/\Tt_1$  versus $\beta$ for an equimolar binary mixture with $\sigma_2/\sigma_1=2$, $m_2/m_1=4$, $\q_1=\q_2=\frac{1}{2}$, $\alpha_{11}=\alpha_{12}=\alpha_{22}=\alpha$, and $\beta_{11}=\beta_{12}=\beta_{22}=\beta$. The values of $\alpha$ are, from bottom to top, $\alpha=1$, $0.9$, $0.8$, $0.7$, $0.6$, and $0.5$. The circles at $\beta=-1$ in panel (c) represent the results obtained in the case of perfectly smooth disks for the same values of $\alpha$.}
\label{fig:Binary}
\end{figure}

\begin{figure*}
\includegraphics[width=\anchotwo\columnwidth]{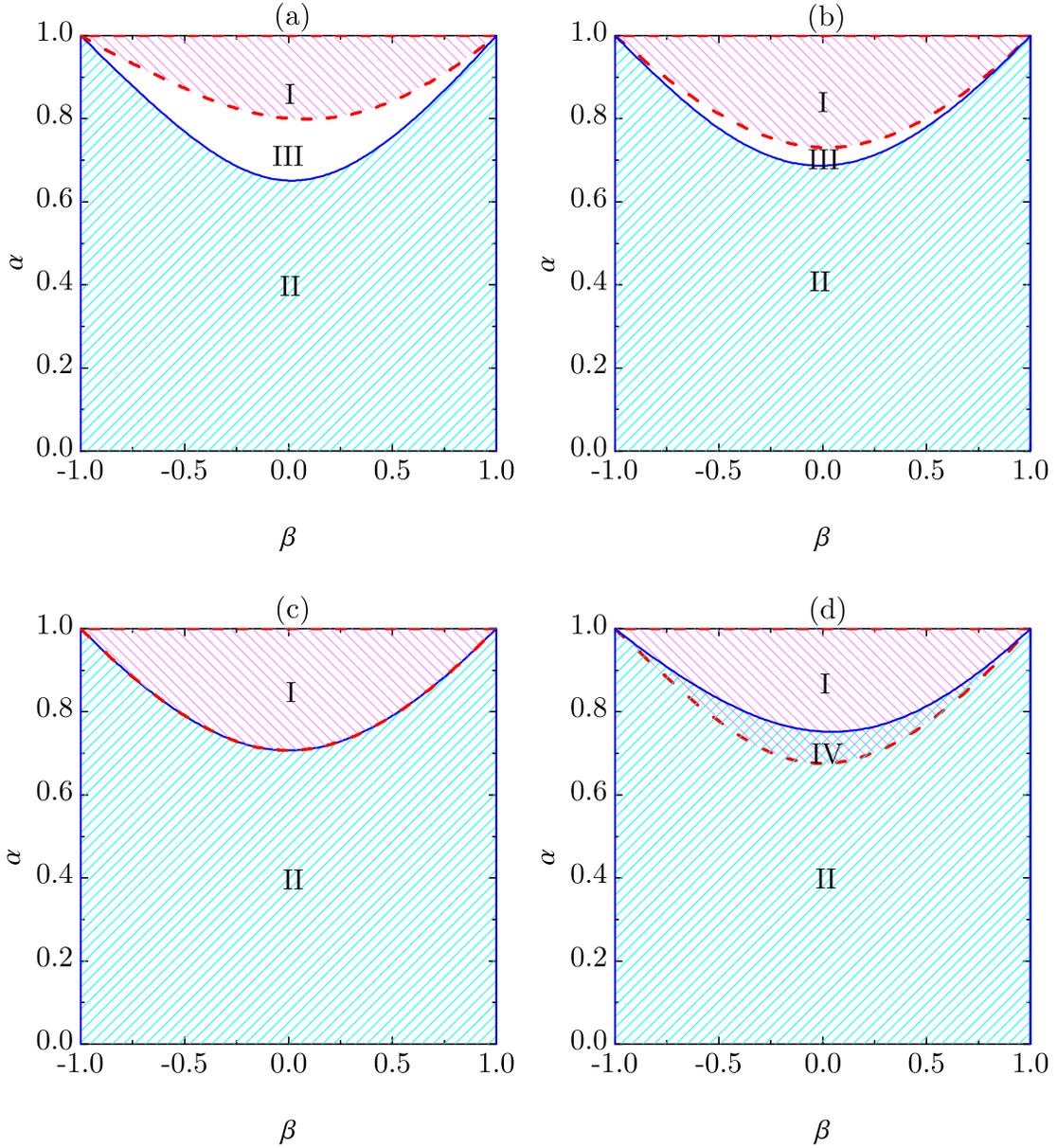}
\caption{Phase diagrams in the case of an equimolar binary mixture ($n_2/n_1=1$) with $\sigma_2/\sigma_1=2$ and (a) $m_2/m_1=4$, (b) $m_2/m_1=2$, (c) $m_2/m_1=1.56541$, and (d) $m_2/m_1=1$.  In regions I and II, one has $T_i^{\text{rot}}<T_i^{\text{tr}}$ and $T_i^{\text{rot}}>T_i^{\text{tr}}$, respectively. In panels (a) and (b), $T_1^{\text{tr,rot}}<T_2^{\text{tr,rot}}$ in the whole plane and  $T_{1,2}^{\text{rot}}\lessgtr T_{1,2}^{\text{tr}}$ in region III. On the other hand, in panel (d), $T_1^{\text{tr,rot}}>T_2^{\text{tr,rot}}$ in the whole plane and  $T_{1,2}^{\text{rot}}\gtrless T_{1,2}^{\text{tr}}$ in region IV. Finally, $\Tt_1=\Tt_2$ and $\Tr_1=\Tr_2$ (mimicry effect) in panel (c). }
\label{fig:phase_diag}
\end{figure*}

In the case of a binary mixture, the three independent temperature ratios ($\Tr_1/\Tt_1$, $\Tt_2/\Tt_1$, and $\Tr_2/\Tr_1$) depend on $12$ free parameters.
As an illustration, let us consider an equimolar mixture where all the disks are uniformly solid  and are made of the same material, the size of the disks of one component being twice that of  the other component. More specifically, $n_2/n_1=1$, $\esn_{11}=\esn_{12}=\esn_{22}=\esn$, $\est_{11}=\est_{12}=\est_{22}=\est$, $\q_1=\q_2=\frac{1}{2}$, $\ds_2/\ds_1=2$, and $m_2/m_1=4$. Moreover, a dilute granular gas is considered ($\phi\ll1$), so that $\chi_{\al\be}\approx 1$. Thus, only the parameters $\esn$ and $\est$ remain free.

Figure \ref{fig:Binary} shows the three independent temperature ratios as functions of the roughness parameter $\est$ for several characteristic values of the inelasticity parameter $\esn$. The rotational-translational temperature ratio $\Tr_1/\Tt_1$ has a behavior qualitatively similar to
that of the monodisperse case (see Fig.\ \ref{fig:Contour}): $\Tr_1/\Tt_1<1$ if $\alpha$ is larger than a certain threshold value ($\alpha=0.651$ in this case) and $\beta$ belongs to a certain $\alpha$-dependent interval around $\beta\approx 0$, whereas $\Tr_1/\Tt_1>1$ otherwise. Moreover,  in the quasismooth limit  $\est\to -1$,  $\Tr_1/\Tt_1$ diverges for inelastic particles ($\esn<1$), while it vanishes for elastic particles ($\esn=1$).
As for the component-component temperature ratios, one has $\Tr_2/\Tr_1>1$ and $\Tt_2/\Tt_1>1$, i.e., the larger disks have larger temperatures than the smaller disks. Additionally, the singularity of $\Tr_1/\Tt_1$ in the limit $\est\to -1$ has a reflection in the rotational-rotational ratio: either $\Tr_2/\Tr_1$ converges to a finite value or it diverges, depending on whether $\esn=1$ or $\esn<1$, respectively. While the ratio $\Tt_2/\Tt_1$ of translational temperatures  remains finite, the huge disparity between the rotational and translational temperatures of both components in the quasismooth limit (if $\esn<1$) has a non-negligible effect on  $\Tt_2/\Tt_1$:  it tends to a value  higher than the one directly obtained in the case of perfectly smooth spheres.
Therefore, a tiny amount of roughness has  dramatic effects on the temperature ratio $\Tt_2/\Tt_1$, producing an enhancement of non-equipartition.

It is interesting to compare the results displayed in Fig.\ \ref{fig:Binary} with those of Fig.\ 2 of Ref.\ \cite{SKG10} for the counterpart case of spheres (i.e., $n_2/n_1=1$, $\esn_{11}=\esn_{12}=\esn_{22}=\esn$, $\est_{11}=\est_{12}=\est_{22}=\est$, $\q_1=\q_2=\frac{2}{5}$, $\ds_2/\ds_1=2$, and $m_2/m_1=8$).
It turns out that, whereas the rotational-translational nonequipartition is stronger in disks than in spheres, the opposite happens with the component-component nonequipartition. For instance, at $\alpha=0.5$ and $\beta=0$ one has $\Tr_1/\Tt_1=1.56$ ($1.36$), $\Tr_2/\Tr_1=4.01$ ($5.25$), and $\Tt_2/\Tt_1=1.75$ ($2.49$) for disks (spheres).

Figure \ref{fig:phase_diag}(a) displays the phase diagram for the two rotational-translational temperature ratios $\Tr_i/\Tt_i$ corresponding to the parameters of Fig.\ \ref{fig:Binary}. The solid and dashed lines represent the loci $\Tr_1/\Tt_1=1$ and $\Tr_2/\Tt_2=1$, respectively. As a consequence, $\Tr_i/\Tt_i<1$ in region I, while $\Tr_i/\Tt_i>1$ in region II. In the intermediate region III, $\Tr_1/\Tt_1<1$ but $\Tr_2/\Tt_2>1$. Apart from that, $\Tr_2/\Tr_1>1$ and $\Tt_2/\Tt_1>1$ in the whole plane, as said before. The same qualitative picture is present if the mass ratio is reduced to $m_2/m_1=2$ (so that $m_2/\sigma_2^2=\frac{1}{2}m_1/\sigma_1^2$), as shown in Fig.\ \ref{fig:phase_diag}(b), except that the loci $\Tr_1/\Tt_1=1$ and $\Tr_2/\Tt_2=1$ approach to each other and thus region III has shrunk with respect to the case of Fig.\ \ref{fig:phase_diag}(a). The situation is reversed in the case of Fig.\ \ref{fig:phase_diag}(d), where $m_2/m_1=1$ (so that $m_2/\sigma_2^2=\frac{1}{4}m_1/\sigma_1^2$). In that case, the locus $\Tr_1/\Tt_1=1$ lies below the locus $\Tr_2/\Tt_2=1$, so that region III has been replaced by region IV, where $\Tr_1/\Tt_1>1$ but $\Tr_2/\Tt_2<1$. In addition, $\Tr_2/\Tr_1<1$ and $\Tt_2/\Tt_1<1$ in the whole plane, i.e., the larger disks have now a smaller temperature. This qualitative change with respect to the cases of Figs.\ \ref{fig:phase_diag}(a) and \ref{fig:phase_diag}(b) is a consequence of the competition between size and mass in the collision frequencies [cf.\ Eq.\ \eqref{56b}]. The transition takes place at $m_2/m_1=1.56541$ (i.e., $m_2/\sigma_2^2\simeq 0.39m_1/\sigma_1^2$), as shown in Fig.\ \ref{fig:phase_diag}(c).
Here, not only the two loci $\Tr_1/\Tt_1=1$ and $\Tr_2/\Tt_2=1$ collapse into a single one (actually, the same as shown in Fig.\ \ref{fig:Contour} for a monodisperse system), but also $\Tr_2/\Tr_1=1$ and $\Tt_2/\Tt_1=1$ in the whole plane. Thus, from the point of view of the mean kinetic energies, the bidisperse gas becomes indistinguishable from a monodisperse gas. This is an example of the mimicry effect further  discussed in Sec.\ \ref{sec7}.

\section{Mimicry effect in the homogeneous cooling state}
\label{sec7}

Imagine a \emph{monodisperse} granular gas (denoted by the label $i=1$) in the HCS, so that its temperature ratio $\Tr_1/\Tt_1$ is the one described in Sec.\ \ref{subsec:Mono}.
Then, we generate a polydisperse gas by adding $\NN-1$ components with the same coefficients of restitution and reduced moments of inertia as the original component $1$, i.e.,
$\alpha_{ij}=\alpha_{11}$, $\beta_{ij}=\beta_{11}$, and $\kappa_i=\kappa_1$.  In general, the addition of the $\NN-1$ extra components produces a new HCS where $\Tr_1/\Tt_1$ is no longer that of a monodisperse gas and, moreover, each component has a different rotational and translational temperature. For instance, this is the situation illustrated  in Figs.\ \ref{fig:Binary}, \ref{fig:phase_diag}(a), \ref{fig:phase_diag}(b), and \ref{fig:phase_diag}(d) for a bidisperse system.

The interesting question is, can we fine-tune the composition, masses, and sizes of the ``invader'' components, so that $\Tr_1/\Tt_1$ is unaltered and $\Tt_i=\Tt_1$, $\Tr_i=\Tr_1$? If so, one can say that a ``mimicry'' effect is present since the $\NN-1$ new components \emph{mimic} the mean kinetic energies of the host gas. To explore that possibility, let us set $\Tt_i=\Tt_1$ and  $\Tr_i=\Tr_1$ in the expressions of $\zt_{ij}$ and $\zr_{ij}$ given in Table \ref{table:2}. This results in
\begin{equation}
\zt_{ij}=\zt_{11}X_{ij},\quad \zr_{ij}=\zr_{11}X_{ij},
\end{equation}
where
\begin{equation}
X_{ij}\equiv\frac{\nu_{ij}}{\nu_{11}}\frac{2m_j}{m_i+m_j}
=\frac{\chi_{ij}n_j\sigma_{ij}}{\chi_{11}n_1\sigma_1}\sqrt{\frac{2m_1m_j}{m_i(m_i+m_j)}}.
\end{equation}
The key point is that the quantities $X_{ij}$ are the same in $\zt_{ij}$ and $\zr_{ij}$. From Eqs.\ \eqref{107}, one has
\begin{equation}
\zt_{i}=\zt_{11}X_{i},\quad \zr_{i}=\zr_{11}X_{i},
\end{equation}
where $X_{i}\equiv\sum_{j=1}^\NN X_{ij}$.
The HCS condition \eqref{59c} implies $\zt_{11}=\zr_{11}$, whose solution gives the ratio $\Tr_1/\Tt_1$ already analyzed in Sec.\ \ref{subsec:Mono}. Next,
Eq.\ \eqref{59a} is equivalent to
\begin{equation}
\label{Xi}
X_1=X_2=\cdots=X_\NN.
\end{equation}
For simplicity, let us assume that the total packing fraction is low enough to make $\chi_{ij}\to 1$. Thus, Eq.\ \eqref{Xi} makes a set of $\NN-1$ constraints on the $3(\NN-1)$ ratios $n_i/n_1$, $\sigma_i/\sigma_1$, and $m_i/m_1$ for $i=2,\ldots,\NN$. In particular, if we freely choose the $2(\NN-1)$ ratios $n_i/n_1$ and $\sigma_i/\sigma_1$, the solution to Eq.\ \eqref{Xi} gives the values of the $\NN-1$ mass ratios $m_i/m_1$ such that the mimicry effect occurs. Without loss of generality, we can assume $n_1\geq n_2\geq \cdots\geq n_\NN$.

In general, the set \eqref{Xi} needs to be solved numerically, but an analytic solution is possible if the intruders have sizes and masses close to those of the host disks. By writing $\sigma_i=\sigma_1(1+\delta\sigma_i^*)$ and $m_i=m_1(1+\delta m_i^*)$, and neglecting terms nonlinear in $\delta\sigma_i^*$ and $\delta m_i^*$, it is straightforward to obtain
\begin{equation}
X_{ij}=\frac{n_j}{n_1}\left(1+\frac{\delta\sigma_i^*+\delta\sigma_j^*}{2}+\frac{\delta m_j^*-3\delta m_i^*}{4}\right),
\end{equation}
\begin{equation}
X_i=\frac{n}{4n_1}\left(2\delta\sigma_i^*-3\delta m_i^*\right)+Y,
\end{equation}
where  the quantity $Y\equiv \sum_{j=1}^\NN (n_j/n_1)(1+\delta\sigma_j^*/2+\delta m_j^*/4)$ is common for all the components. Therefore, Eq.\ \eqref{Xi} yields $2\delta\sigma_i^*-3\delta m_i^*=0$ for $i=2,\ldots,\NN$ or, equivalently
\begin{equation}
\label{slope}
\frac{m_i}{m_1}\approx \frac{1+2\sigma_i/\sigma_1}{3} \quad (\sigma_i\approx\sigma_1),
\end{equation}
regardless of $n_i/n_1$.
Since it has been assumed that $\sigma_i\approx\sigma_1$, and thus all the components are similar, it is convenient to convert Eq.\ \eqref{slope} into  a form independent of the choice for the reference component. This is accomplished by replacing $m_i\propto 1+2\sigma_i/\sigma_1$ by
$m_i\propto 1+2\sigma_i/\langle\ds\rangle$, where $\langle\ds\rangle=n^{-1}\sum_{j=1}^\NN n_j\ds_j$ is the mean diameter. Therefore,
\begin{equation}
\label{slope2}
\frac{m_i}{m_1}\approx \frac{1+2\sigma_i/\langle \sigma\rangle}{1+2\sigma_1/\langle \sigma\rangle} .
\end{equation}
As will be seen in Secs.\ \ref{subsec:bin} to \ref{subsec:conti}, Eq.\ \eqref{slope2} turns out to be an excellent approximation.

\subsection{Binary mixture}
\label{subsec:bin}

\begin{figure}
\includegraphics[width=\ancho\columnwidth]{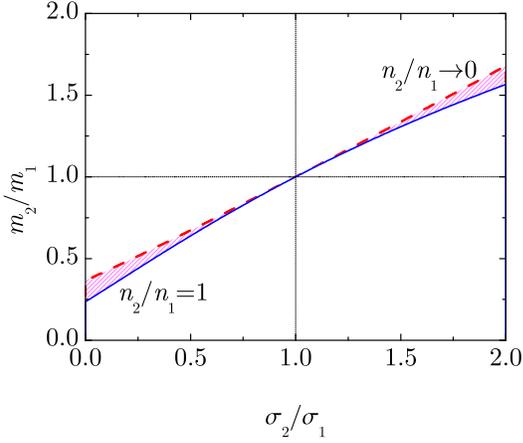}
\caption{The hatched region represents the
values of $m_2/m_1$ and $\sigma_2/\sigma_1$ where mimicry is possible  in a binary mixture [see Eq.\ \eqref{115}]. The boundaries of the region correspond to the  extreme compositions $n_2/n_1\to 0$ and $n_2/n_1=1$.  }
\label{fig:bin}
\end{figure}

In the case of a binary mixture ($\NN=2$), the condition $X_2=X_1$ becomes
\begin{equation}
\frac{n_2}{n_1}=\frac{\frac{\ds_{12}}{\ds_1}\sqrt{\frac{m_1}{m_2}}-\sqrt{\frac{m_1+m_2}{2m_1}}}{\frac{\ds_{12}}{\ds_1}\sqrt{\frac{m_2}{m_1}}-\frac{\ds_{2}}{\ds_1}\sqrt{\frac{m_1+m_2}{2m_2}}}.
\label{115}
\end{equation}
Thus, if $n_2/n_1$ and $\sigma_2/\sigma_1$ are freely chosen, Eq.\ \eqref{115} gives the value of $m_2/m_1$ corresponding to the mimicry effect. In particular, in the tracer limit  $n_2/n_1\to 0$ the solution is
\begin{equation}
\label{tracer}
\frac{m_2}{m_1}=\sqrt{\frac{3}{4}+\frac{\sigma_2}{\sigma_1}+\frac{\sigma_2^2}{2\sigma_1^2}}-\frac{1}{2}\quad \left(\frac{n_2}{n_1}\to 0\right).
\end{equation}
In this tracer limit, $\langle\sigma\rangle=\ds_1$, so that Eqs.\ \eqref{slope} and \eqref{slope2} are identical. Interestingly, Eq.\ \eqref{tracer} deviates very little from Eq.\ \eqref{slope}, the maximum relative deviation (less that $10\%$) taking place in the limit $\sigma_2/\sigma_1\to 0$.

Figure \ref{fig:bin} plots the mass ratio  $m_2/m_1$ as a function of the size ratio $\sigma_2/\sigma_1$ for $n_2/n_1\to 0$  and $n_2/n_1=1$. The curves corresponding to intermediate values of $n_2/n_1$ lie in the hatched region comprised by those two curves. For instance, if $n_2/n_1=1$ and $\ds_2/\ds_1=2$, then $m_2/m_1=1.56541$, and this is the case considered in Fig.\ \ref{fig:phase_diag}(c).
The slope of the curves $n_2/n_1=\text{const}$   at $\sigma_2/\sigma_1=1$ is $\frac{2}{3}$ with independence of the value of $n_2/n_1$, in agreement with Eq.\ \eqref{slope}. In fact, the deviations from the linear behavior given by Eq.\ \eqref{slope} are small in the tracer case ($n_2/n_1\to 0$), as said before, and not particularly large in the equimolar case ($n_2/n_1=1$). On the other hand, if $n_2/n_1=1$, Eq.\ \eqref{slope2} yields the nonlinear approximation $m_2/m_1=(1+5\ds_2/\ds_1)/(5+\ds_2/\ds_1)$, which performs excellently well, with a maximum deviation $0.036$ at $\sigma_2/\sigma_1=0$.

{}From Fig.\ \ref{fig:bin} we can observe that $m_2/m_1>(\ds_2/\ds_1)^2$ and $m_2/m_1<(\ds_2/\ds_1)^2$ if $\ds_2/\ds_1<1$ and $\ds_2/\ds_1>1$, respectively. Therefore, a necessary condition for the existence of the mimicry effect is that the smaller disks must have a higher solid density than the larger disks.

\begin{figure}
\includegraphics[width=\ancho\columnwidth]{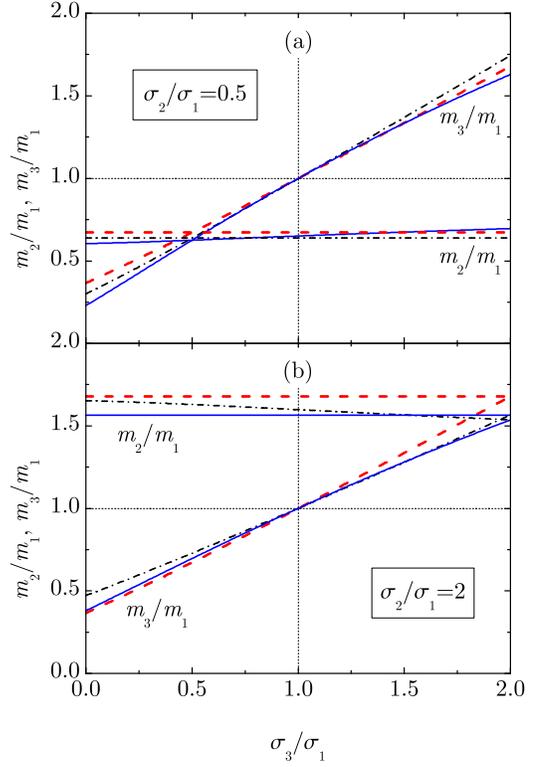}
\caption{Plot of $m_2/m_1$ and $m_3/m_1$ versus $\sigma_3/\sigma_1$ for mimicry in a ternary mixture with (a) $\sigma_2/\sigma_1=0.5$ and (b) $\sigma_2/\sigma_1=2$. The solid line, dash-dotted  line, and dashed line correspond to the compositions $(n_2/n_1,n_3/n_1)=(1,1)$, $(1,0)$, and $(0,0)$, respectively.}
\label{fig:ter}
\end{figure}

\begin{figure}
\includegraphics[width=\ancho\columnwidth]{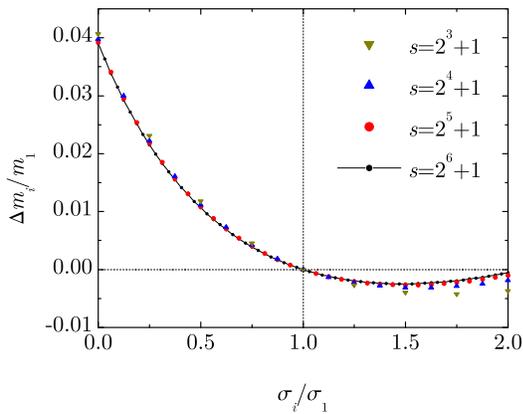}
\caption{Plot of the difference between $m_i/m_1$ and the estimate \eqref{slope} versus $\sigma_i/\sigma_1$ for mimicry in a polydisperse gas described by Eq.\ \eqref{poly}.}
\label{fig:poly}
\end{figure}

\subsection{Ternary mixture}
\label{subsec:ter}
Obviously, the ternary case ($\NN=3$) is more complex than the binary one. Now we have the freedom to choose $n_2/n_1$, $n_3/n_1$, $\ds_2/\ds_1$, and $\ds_3/\ds_1$. Then, $m_2/m_1$ and $m_3/m_1$ are obtained from $X_1=X_2=X_3$.

To be more specific, let us choose three possible compositions: $(n_2/n_1,n_3/n_1)=(1,1)$, $(1,0)$, and $(0,0)$. The first case corresponds to an equimolar ternary mixture, while in the third case the two intruder components $i=2,3$ are tracer particles; in the second case, tracer particles of component $i=3$ are added to an equimolar binary mixture already exhibiting mimicry. Additionally, $\ds_2/\ds_1=0.5$ and $\ds_2/\ds_1=2$ are chosen. For those six systems, Fig.\ \ref{fig:ter} shows $m_2/m_1$ and $m_3/m_1$ as functions of $\ds_3/\ds_1$. From the rough estimate of Eq.\ \eqref{slope}, one obtains $m_2/m_1\approx 0.7$ and $m_2/m_1\approx 1.7$ for $\ds_2/\ds_1=0.5$ and $\ds_2/\ds_1=2$, respectively, with independence of composition and $\ds_3/\ds_1$. A much better prediction for $m_2/m_1$ is obtained from Eq.\ \eqref{slope2}, which yields a maximum deviation of $0.015$ in the case $(n_2/n_1,n_3/n_1)=(1,1)$ and $(\ds_2/\ds_1,\ds_3/\ds_1)=(2,0)$. Moreover, the curves representing $m_3/m_1$ as functions of $\ds_3/\ds_1$ are also roughly similar to the linear behavior \eqref{slope}, but again the approximation \eqref{slope2} is very accurate, with a maximum deviation of $0.047$ taking place at the same state  [$(n_2/n_1,n_3/n_1)=(1,1)$ and $(\ds_2/\ds_1,\ds_3/\ds_1)=(2,0)$] as before.

\subsection{Toward a continuous size distribution}
\label{subsec:conti}
Consider now a polydisperse gas with a continuous size distribution $n(\sigma)$ such that $n(\sigma)d\sigma$ is the number of disks per unit area with a diameter between $\sigma$ and $\sigma+d\sigma$. In that case, Eq.\ \eqref{Xi} becomes
\begin{subequations}
\label{conti}
\begin{equation}
\label{conti_a}
\frac{\partial}{\partial\sigma}X(\sigma)=0,\quad X(\sigma)\equiv \int_0^\infty d\sigma'\,n(\sigma')X(\sigma,\sigma'),
\end{equation}
\begin{equation}
\label{conti_b}
X(\sigma,\sigma')\propto \frac{\sigma+\sigma'}{\sqrt{m(\sigma)}}\sqrt{\frac{m(\sigma')}{m(\sigma)+m(\sigma')}},
\end{equation}
\end{subequations}
where $m(\sigma)$ is the mass of a particle of diameter $\sigma$. Given a certain size distribution $n(\sigma)$, Eq.\ \eqref{conti_a} is an integro-differential equation for $m(\sigma)$ which, in general, can be difficult to solve.

On the other hand, using Eq.\ \eqref{slope} as a starting guess, it is quite possible to solve numerically Eq.\ \eqref{Xi} for a discrete mixture with a large number of components, thus mimicking a continuous distribution \cite{UKAZ09}. As an example, let us take an equimolar mixture ($n_i/n_1=1$) with a number of components $\NN=\text{odd}$ and sizes
\begin{equation}
\label{poly}
\frac{\sigma_i}{\sigma_1}=\begin{cases}
  \displaystyle{2\frac{i-2}{\NN-1}},& \displaystyle{2\leq i\leq \frac{\NN+1}{2}},\\
   \displaystyle{2\frac{i-1}{\NN-1}},& \displaystyle{\frac{\NN+3}{2}\leq i\leq \NN}.
\end{cases}
\end{equation}
Note that $\sigma_1$ coincides with the mean diameter, i.e., $\langle \ds\rangle=\sigma_1$, so that Eqs.\ \eqref{slope} and \eqref{slope2} are fully equivalent.
In the limit $\NN\to\infty$ this discrete mixture becomes a continuous system with a uniform distribution of sizes  between $\sigma=0$ and $\sigma=2\langle\ds\rangle$.

The solution of Eq.\ \eqref{Xi} for the above class of mixtures converges to a mass distribution very close to the simple estimate \eqref{slope}. This is observed in Fig.\ \ref{fig:poly}, which plots the difference $\Delta m_i/m_1=m_i/m_1-(1+2\sigma_i/\sigma_1)/3$ versus $\sigma_i/\sigma_1$ for $\NN=2^q+1$ with $q=3,4,5,6$. As can be observed, the convergence to a continuous curve is quite apparent, the results obtained with $s=2^5+1=33$ being highly consistent with those obtained with $s=2^6+1=65$. Again, the maximum deviation ($\Delta m_i/m_1=0.039$) takes place in the limit $\sigma_i\to 0$.

The mimicry effect described in this section assumes that all the components have  common coefficients of normal and tangential restitution. As an important consequence,  the conditions for mimicry turn out to be independent of the specific values of those coefficients. Of course, this is not the general case. If not all the coefficients of restitution are equal, the conditions for mimicry are obtained by inserting $\Tt_i\to\Tt$ and $\Tr_i\to\Tr$ into the production rates $\zt_{ij}$ and $\zr_{ij}$, and applying Eqs.\ \eqref{59}. This gives the ratio $\theta\equiv \Tr/\Tt$ and provides, in general, $2(\NN-1)$ constraints on the $\NN-1$ density ratios $\{n_i/n_1\}$, the $\NN-1$ size ratios $\{\sigma_i/\sigma_1\}$, the $\NN-1$ mass ratios $\{m_i/m_1\}$, the $\NN$ reduced moments of inertia  $\{\qa\}$, the $\NN(\NN+1)/2$ coefficients of normal restitution $\{\alpha_{ij}\}$, and the $\NN(\NN+1)/2$ coefficients of tangential restitution $\{\beta_{ij}\}$.

\section{Concluding remarks}
\label{sec:conc}
Granular gases of inelastic and rough hard disks have a two-fold importance. On the one hand, they are prototypical models for most of the experimental setups related to granular matter under conditions of rapid flow. On the other hand, they pose an interesting physical problem by its own since, in contrast to the case of  spheres, the two vector subspaces associated with the translational and angular degrees of freedom are mutually orthogonal.

While monodisperse frictional hard-disk systems have been analyzed  by kinetic-theory tools before \cite{JR85a,LHMZ98,HCZHL05,DF17}, the emphasis here has been on the crossed collisional rates of change of energy ($\zabt$ and $\zabr$) for a multicomponent gas.
Starting from the collisional rules \eqref{13b&14}, together with Eq.\ \eqref{15}, the energy production rates can be expressed in a formally exact way in terms of the two-body distribution function $\fab$ [see Eqs.\  \eqref{15c}, \eqref{Z1}, and \eqref{3}]. Next, the original function $\fab$ has been replaced by its pre-collisional orientational average $\ffab$  [see Eq.\ \eqref{III.7}], this assumption being justified if the density and/or the heterogeneities are small. This allows for the expression of the collisional rates of change as combinations of two-body averages, as shown in Table \ref{table:0}. Explicit results as functions of densities, temperatures, and mean angular velocities are then obtained by a maximum-entropy approach [see Eq.\ \eqref{IV.1}], implying molecular chaos, rotational-translational statistical independence, and a Maxwellian translational velocity distribution. The final expressions, summarized in Table \ref{table:2}, represent the primary contribution of this paper.

The most immediate application of the results reported here has been the study of the HCS regime (where all the partial temperatures decay at the same rate), even though the transient regime to the asymptotic state can present interesting and counterintuitive phenomena \cite{PT14,TP14,LVPS17}.
In comparison to the hard-sphere case, it is found that the degree of breakdown of energy equipartition in hard-disk gases has a dual character: disks typically present a stronger rotational-translational nonequipartition but a weaker component-component nonequipartition than spheres.

Special attention has been paid to the mimicry effect. This effect consists in the possibility of adding to a monodisperse gas ($i=1$) an arbitrary number ($\NN-1$) of components with arbitrary concentrations ($n_i$) and arbitrary diameters ($\sigma_i$), but with the same coefficients of restitution ($\alpha_{ij}=\alpha_{11}$, $\beta_{ij}=\beta_{11}$) and reduced moment of inertia ($\kappa_i=\kappa_1$) as in the host system,  in such a way that the translational and rotational temperatures are the same as those of the original monodisperse system (i.e., $\Tat=\Tt_1$, $\Tar=\Tr_1$).
This requires the fine-tuning of the mass ($m_i$) of each invader component as a function of the values of $\{n_j\}$ and $\{\sigma_j\}$, the results being independent of $\alpha_{11}$, $\beta_{11}$, and $\kappa_1$. A simple (but yet rather accurate) coarse-grained recipe turns out to be $m_i\propto 1+2\sigma_i/\langle \sigma\rangle$, so that the mass per unit area $m_i/(\frac{\pi}{4}\sigma_i^2)$ decreases with increasing size. It might seem artificial that all the disks have the same coefficients of restitution and reduced moment of inertia (thus apparently being made of the same material) and yet have different masses per unit area. But, in contrast to the case of spheres, there is a simple possibility of experimental realization by  considering that the disks actually correspond to vertically aligned cylinders with different diameters ($\sigma_i$) and heights ($h_i$) but the same mass per unit volume ($\rho$), so that $m_i=\rho\frac{\pi}{4}\sigma_i^2 h_i$. In that case, the approximate condition $m_i\propto 1+2\sigma_i/\langle \sigma\rangle$ translates into $h_i\propto \sigma_i^{-1}\left(\sigma_i^{-1}+2\langle \sigma\rangle^{-1}\right)$.

Analogously to the case of hard spheres \cite{VLSG17,VLSG17b}, the expressions derived in this work are expected to compare well with computer simulations, and a critical assessment is planned in the near future. In addition, once the energy production rates are known, the study of hard-disk gases driven stochastically \cite{VS15} is straightforward and will also be carried out and compared with simulation. Finally, and more importantly, the results derived here lay the basis for the study of nonuniform situations. Taking the local version of the HCS as the reference state, a Chapman--Enskog method can be followed to derive the Navier--Stokes constitutive equations and analyze the linear stability conditions of the HCS, in analogy with what has recently been done in the case of rough spheres \cite{KSG14,GSK18}.
Moreover, in the case of a binary mixture, the hydrodynamic equations  stemming from the HCS  can be used to study the conditions for segregation under the presence of a thermal gradient \cite{G06a,GMV13}.

\begin{acknowledgments}
Financial support from the Ministerio de Econom\'ia y Competitividad (Spain) through Grant No.\ FIS2016-76359-P
and from the Junta de Extremadura
(Spain) through Grant No.\ GR18079, both partially financed by ``Fondo Europeo de Desarrollo Regional'' funds, is gratefully acknowledged.
\end{acknowledgments}

\bibliography{D:/Dropbox/Mis_Dropcumentos/bib_files/Granular}

\end{document}